# PHYSICAL PROPERTY CHARACTERIZATION OF BULK $MgB_2$ SUPERCONDUCTOR

V.P.S. Awana[$], Arpita Vajpayee, Monika Mudgel, and H. Kishan

Superconductivity and Cryogenics Division, National Physical Laboratory, Dr. K.S. Krishnan Marg, New Delhi-110012, India

V. Ganesan and A.M. Awasthi

UGC-DAE Consortium for Scientific Research, University Campus, Khandwa Road, Indore-452017, India

G. L. Bhalla

Deptartment of Physics and Astrophysics, University of Delhi, New Delhi-110007, India

We report synthesis, structure/micro-structure, resistivity under magnetic field [$\rho(T)H$], Raman spectra, thermoelectric power $S(T)$, thermal conductivity $\kappa(T)$, and magnetization of ambient pressure argon annealed polycrystalline bulk samples of $MgB_2$, processed under identical conditions. The compound crystallizes in hexagonal structure with space group *P*6/*mmm*. Transmission electron microscopy (*TEM*) reveals electron micrographs showing various types of defect features along with the presence of 3-4nm thick amorphous layers forming the grain boundaries of otherwise crystalline $MgB_2$. Raman spectra of the compound at room temperature exhibited characteristic phonon peak at 600 $cm^{-1}$. Superconductivity is observed at 37.2K by magnetic susceptibility $\chi(T)$, resistivity $\rho(T)$, thermoelectric power $S(T)$, and thermal conductivity $\kappa(T)$ measurements. The power law fitting of $\rho(T)$ give rise to Debye temperature ($\Theta_D$) at 1400K which is found consistent with the theoretical fitting of $S(T)$, exhibiting $\Theta_D$ of 1410K and carrier density of $3.81\text{x}\ 10^{28}/m^3$. Thermal conductivity $\kappa(T)$ shows a jump at 38K, i.e., at $T_c$, which was missing in some earlier reports. Critical current density ($J_c$) of up to $10^5$ A/$cm^2$ in 1-2T (Tesla) fields at temperatures ($T$) of up to 10K is seen from magnetization measurements. The irreversibility field, defined as the field related to merging of $M(H)$ loops is found to be 78, 68 and 42 kOe at 4, 10 and 20K respectively. The superconducting performance parameters viz. irreversibility field ($H_{irr}$) and critical current density $J_c(H)$ of the studied $MgB_2$ are improved profoundly with addition of *nano*-SiC and *nano*-Diamond.

The physical property parameters measured for polycrystalline $MgB_2$ are compared with earlier reports and a consolidated insight of various physical properties is presented.

Key Words: $MgB_2$, Phase formation, Magneto-transport, Thermoelectric power, Thermal conductivity, Critical current density, and Magnetization.

$: Corresponding author for further queries: awana@mail.nplindia.ernet.in

## 1. INTRODUCTION

Discovery of superconductivity in $MgB_2$ with the critical temperature of 39 K [1] had been of prime importance for theorists and experimentalists alike. Basically the higher critical temperature ($T_c$) of $MgB_2$ in comparison to other superconductors, except the high $T_c$ cuprates (i.e., HTSc), was particularly intriguing as to whether the usual electron – phonon interactions could explain such a high $T_c$. By now some consensus has been reached regarding the pairing mechanism and electronic structure of this compound, see review articles [2-4], and references there in. In fact evidence of isotope effect yielded a clear indication that phonons play an important role in pairing mechanism in this compound [5,6]. On the other hand the 39 K superconductivity of $MgB_2$ could be explained by its multiple superconducting gaps structure [7,8]. For more clarifications and relevant references in regards to the two band superconductivity and inter band scattering etc, see the very recent review in ref. [9].

To experimentalists, the relatively higher $T_c$ [1], simpler structure with lower anisotropy [10] and nearly transparent grain boundaries [11] promise huge scope for practical applications of $MgB_2$. The material has already been studied rigorously in terms of its crystal structure, thermal and electrical conduction [12-14], specific heat [15, 16], isotope effect [5,6] and doping effects etc [17-19]. Its $T_c$ of about 40 K and better material properties respectively in comparison to so called inter-metallic *BCS* type superconductors and *HTSc* cuprates provide $MgB_2$ with an edge over other superconductors. Further, relatively higher coherence length of about 5 nm of MgB2, in comparison to HTSc compounds, permits the use of nano particles of various compounds viz., SiC [20, 21], carbon-*nano*tubes [22] and *nano*-diamonds [23], etc., introduced as additives, to act as effective pinning centers and thereby enhance the critical current ($J_c$) for the $MgB_2$ superconductor. Needless to say, $MgB_2$ is unique in many ways regarding its fundamental importance in the field of superconductivity.

In the present article we have studied various physical properties of polycrystalline bulk $MgB_2$ and have compared them with previously reported data in the literature. In particular, we focus on synthesis, structure/micro-structure, resistivity under magnetic field [$\rho(T)H$], Raman spectroscopy, thermoelectric power $S(T)$, thermal conductivity $\kappa(T)$ and magnetization of ambient pressure argon annealed (850 $^0$C) polycrystalline bulk sample. Various routes of synthesis including vacuum/Argon annealing at various temperature are applied to find the optimum-heating schedule. For technical applicability, *nano*-additives like *nano*-SiC and *nano*-Diamonds are added to the title compound and its enhanced performance is presented in short at the end of the paper.

## 2. EXPERIMENTAL

Various polycrystalline $MgB_2$ samples are synthesized by solid-state route using ingredients of Mg and B. The Mg powder used is from *Reidel-de-Haen* of assay 99%, insoluble in HCl and with Fe impurity of less than 0.05%. B powder is amorphous and *Fluka* make of assay 95-97%. For high superconducting performance of $MgB_2$ the *nano*-additives used are *nano*-SiC (size 10-20 nm) and *nano*-diamond (size 7-10 nm). For individual sample, well-mixed and pelletized rectangular $MgB_2$ pellet is put in an Alumina boat placed inside a soft Fe-tube. The encapsulated system is heated in the temperature range from 800 $^0$C to 900 $^0$C for one to three hours in flow of Argon gas at ambient pressure and subsequently allowed to cool to room temperature in same atmosphere. The Fe-encapsulation is not kept in direct touch of the raw $MgB_2$ pellet and its both ends are open for the continuous passage of Argon gas. The resultant sample is a bulk polycrystalline black compound. We also synthesized $MgB_2$ by vacuum annealing method. Pure phase $MgB_2$ can be synthesized in oxygen free environment from 700 $^0$C to 1400 $^0$C, respectively in vacuum ($10^{-5}$ torr) to high pressure of Argon gas [24, 25]. In this case the Fe tube, which contains the raw MgB2 pellet inside, was sealed inside a quartz tube at high vacuum of $10^{-5}$ Torr. The encapsulated raw $MgB_2$ pellet is than heated at desired heating temperatures (750 $^0$C) with a hold time (3 hours) and is finally quenched in liquid nitrogen ($LN_2$) [26,27]. However we realized that the vacuum annealing method results in highly porous samples [28], which in turn adversely affect the transport properties like electrical and thermal conduction, in particular. The effects of porosity, crystallization and alignment of grains on various physical properties of $MgB_2$ are previously discussed in a topical review by John M Rowell [29].

The x-ray diffraction pattern of the compound is recorded with a Diffractometer using $CuK_\alpha$ radiation. The scanning electron Microscopy (*SEM*) studies are carried out on these samples using a Leo 440 (Oxford Microscopy: UK) instrument. *TEM* is performed using *Tecnie G2-20* equipped with $LaB_6$ filament and operating at 200kV. The electron micrographs are recorded using CCD camera attached to the *TEM*. Raman measurements were performed on a dispersive single Horiba–Jobin–Yvon Hr-800 mono-chromator coupled to a charge couple device. The 488 line of an argon ion laser was used as a probe beam that is focused on to a ~2 µm spot. The power was kept to a minimum of ~2 mW at the sample, and all the measurements were carried out in a back scattering geometry with detection in the un-polarized mode. Resistivity measurements are carried out by four-probe technique under applied fields of up to 80 kOe. Thermoelectric power (*TEP*) measurements are carried out by dc differential technique over a temperature range of 5 – 300K, using a homemade set-up. Temperature gradient of ~1K is maintained throughout the *TEP* measurements. Thermal conductivity data are obtained by conventional steady state method on a polycrystalline pellet. The dimensions of the measured pellet are: length (5.36mm), width (5.34mm) and thickness (4.5mm). Temperature gradient ($\Delta T$) of 0.2K to 0.5K is developed by excitation of a strain-gauge resistor (thermally attached to flat sample surface), at constant differential heater power (3-10mW). Sample's operational temperature $T$ over its second parallel face is maintained within 5mK using Lakeshore DRC-93CA controller, and the fluctuation of $\Delta T$ kept within ±1%. Magnetization measurements are carried out with a Quantum-Design 14 Tesla Physical Property measurement System (*PPMS)* having Vibrating Sample Magnetometer (*VSM*) attachment.

## 3. RESULTS AND DISCUSSION

### 3A. Structure and Microstructure

Figure1 depicts the room temperature x-ray diffraction (*XRD*) pattern of variously synthesized $MgB_2$ samples. The indexing of respective *XRD* peaks corresponds to known hexagonal Bravais lattice. Seemingly all the samples are near single phase till heating temperature of 850 $^{0}$C and hold time of 1-3 hours. At 900 $^{0}$C and above the resultant compound is clearly multi phase. A careful look at the observed peak intensities implies that though 800 $^{0}$C heated samples are all nearly phase pure, the 850 $^{0}$C – 2 hours annealed sample gives the best fit to the observed intensities, see Fig. 2. Except for a weak reflection at $2\theta = 63^{0}$, corresponding to pure MgO [2,17,25-28], the rest of the Bragg reflections are characteristic of the hexagonal $MgB_2$ structure. MgO peak is marked in the pattern in Fig. 2. The structure of $MgB_2$ belongs to space group *P6/mmm*. The asymmetric unit of the structure consists of Mg at (0, 0, 0) and B at (1/3, 2/3, 1/2). Preliminary Rietveld refinement is carried out using the program Fullprof [30]. The impurity peak is excluded during the refinement. The occupancy parameters of the various atoms and their positions are fixed at their nominal values. Details of the refinement procedure along with anomalous lattice parameters variation around $T_c$ will be reported elsewhere [31]. The lattice parameters are $a$ = 3.08 Å, and $c$ = 3.53 Å, with $c/a$ ~ 1.14. The lattice parameters are though close to our vacuum annealed $MgB_2$ [27], the presence of MgO is comparatively less in Argon annealed sample, as seen by *XRD* pattern (Fig.2). As far as the presence of minute amount of MgO is concerned, the same is seen in earlier reports as well [2,15,25-28, 32-37], but not always marked on the respective *XRD* pattern.

Scanning electron microscope (*SEM*) pictures of present sample are shown in Fig. 3(a) and 3(b) with two different magnifications. Nearly homogenous distribution of crystallites can be seen in *SEM* picture in Fig. 3(a). The average grain shape is like platelets with size 2-5 $\mu m$. The shape and size of observed grains for present $MgB_2$ are in general agreement with the reported literature on this compound [32-34].

Out of bulk pellet of $MgB_2$, thin samples for *TEM* observation were prepared using ion-beam polishing technique. For ion polishing, 3 kV argon ion beam was used at about 3$^{o}$ grazing incidence. The representative defect microstructures observed during *TEM* analysis are shown in the micrographs of Figs. 4 and 5.

Figs. 4(a) and (b) respectively represent bright-field and corresponding dark-field images. These micrographs show the presence of various types of defect features in the $MgB_2$ matrix. Also seen are nearly spherical inclusions of sizes 4-15*nm* in an otherwise crystalline matrix of $MgB_2$. These inclusions are analyzed to be MgO [32]. The presence of *nano*-particle inclusions of MgO can be seen more clearly in the *TEM* micrograph shown in Fig.4(c), which was taken at different tilt and at higher magnification. Other defect feature present in the matrix is the stacking fault ribbons with partial dislocations.

Their presence can be seen in micrographs 4(a) and (b) by the fault fringes confined between two dislocation lines.

Other micro-structural features, which were seen during *TEM* analysis, are the presence of 3-4*nm* thick amorphous layers between the grain boundaries of otherwise crystalline $MgB_2$. These thin amorphous layers at the grain boundary are shown in Fig. 5 (a) and (b). At some places these layers were found to contain 2-3*nm* precipitates. The width of the layer was too narrow for the selected area electron diffraction using even the smallest aperture available. However, the amorphous nature of the layers was convincingly established by large angle tilting of the sample that did not result in any change in the image contrast of the layer. These defect features may be responsible for pinning induced enhancement of critical current. The diffused area is seemingly responsible for the improved grains connectivity and their transparency to the current transport.

### 3B. Raman studies

Room temperature Raman spectrum of the optimized (850 $^0$C – 2 hours) pristine $MgB_2$ sample is depicted in Fig. 6. The phonon peak occurs at 600 cm$^{-1}$, this is in confirmation with various earlier reports on this compound [35-37]. The 600 cm$^{-1}$ is the characteristic e-ph (electron-phonon) coupling peak for superconducting $MgB_2$. This arises from $E_{2g}$ phonon mode, being ascribed to the in-plane B bond stretching. Any disorder in terms of onsite substitution in $MgB_2$ results in weakening of e-ph coupling and a shift in peak position to higher energies. The 600 cm$^{-1}$ e-ph peak in $MgB_2$ is unusual for $AlB_2$ class (hexagonal *P*6/*mmm*) materials. In fact for $AlB_2$ the e-ph peak occurs at ~ 980 cm$^{-1}$ [37]. Strong electron-phonon coupling due to softening of e-ph modes is the main reason behind 39K superconductivity of $MgB_2$. As mentioned in introduction the phonons play an important role in pairing mechanism in this compound [5,6], and the 39 K superconductivity could be just at the BCS strong coupling limit. An early indication of the strong phonon contribution in $MgB_2$ can be presumed on the basis of its unusually stretched *c*-lattice parameter. The *c*/*a* values are 1.14 and 1.08 respectively for $MgB_2$ and $AlB_2$. This clearly indicates that $MgB_2$ is stretched in *c* direction in comparison to $AlB_2$. Worth mentioning is the fact that there is no much difference in the ionic sizes of Mg and Al. Hence the stretching of $MgB_2$ lattice in comparison to $AlB_2$ can not be explained simply on the basis of crystal chemistry and the electronic changes need to be probed. This explains the unusual e-ph peak position (600 $cm^{-1}$) for $MgB_2$ in comparison to at 980 $cm^{-1}$ for $AlB_2$. The 600 $cm^{-1}$ e-ph peak of $MgB_2$ shifts to higher wave number side for $Mg_{0.6}Al_{0.4}B_2$, see upper plot in Fig. 6. A similar case is observed for $MgB_{2-x}C_x$ as well [37]. Seemingly the phonon peak at 600 $cm^{-1}$ is seen only in superconducting $MgB_2$, which moves towards higher wave number side accompanied by deterioration in the superconductivity.

### 3C. Resistivity Analysis

The resistivity $\rho(T)$ plot of polycrystalline $MgB_2$ is shown in the Fig. 7. The critical temperature $T_c$ and room temperature resistivity ($\rho^{300K}$) are found to be 37K and 78μΩ.cm respectively. The superconducting transition temperature ($T_c$) is defined at temperatures where $\rho \rightarrow 0$. For very phase pure polycrystalline $MgB_2$, the $\rho^{300K}$ value in literature is reported around10-100μΩ-cm [5, 33, 38-40]. In fact in polycrystalline samples, the connectivity of grains affects the conduction in a major way [29]. Moreover the porous nature and low theoretical density (~ 1.6 gram/$cm^3$) of $MgB_2$ makes this problem more serious. Though the Ar-annealed $MgB_2$ samples are comparatively less porous than the vacuum annealed ones [28], the porous regions are still seen clearly in *SEM* pictures, see Fig. 3. Though there can be various other causes behind less connectivity of grains in polycrystalline $MgB_2$ such as the presence of insulating MgO (Fig. 4), we feel the main cause is its porous nature. Following the suggestion of Rowell in ref. 29, the connectivity of grains can be estimated by knowing the value of $\rho^{300K}$ - $\rho^{40K}$, which is ~ 44μΩ.cm in present sample. The same value is ~ 7-8 μΩ.cm for thin films [29]. This roughly shows that the grains connectivity of presently studied polycrystalline $MgB_2$ sample is 6 times less than as for any well connected $MgB_2$. Hence the corrected resistivity, which is the true resistivity within the grains themselves, is 78/6 = 13μΩ.cm in present case. The conduction of carriers gets deteriorated when their mean free path becomes comparable to the disorder within or outside the grains. In case of $MgB_2$ the disorder with in grains could occur due to varying perfection, or the degree of crystalline disorder in differently synthesized samples. Outside grains disorder mainly comes from the embedded impurities or the porous regions. As far as grain boundaries are concerned, the same are though reported to be transparent to the current [11], their actual full transparency to conduction in variously processed samples is debatable [29]. A careful comparison of the resistivity values of irradiation induced disordered superconductors by J.M. Rowell, showed that outside grains impurities/porous regions are mainly responsible for the widely varying conductivity of the $MgB_2$ superconductor [29]. For most of studied bulk polycrystalline samples [35,38-40] along with the present one, the problem of connectivity is more serious, than better-connected most of $MgB_2$ thin films [29].

In normal state i.e., above $T_c$ onset, the compound is metallic with residual resistivity ratio ($\rho^{300K}/\rho^{onset}$) of around 2.25, which is generally defined by *RRR* (residual resistivity ratio). In literature the *RRR* is reported to be up to 20 for very pure disorder free dense samples [5]. For disordered samples the *RRR* comes down rapidly, for example *RRR* is only 1.5 for $MgB_{2-x}C_x$ samples [39]. The *RRR* of polycrystalline bulk $MgB_2$ in ref. 33 is 3.0, which is close to the present value of 2.25. Both slightly higher $\rho^{300K}$ and low *RRR* values demonstrate that studied $MgB_2$ sample is disordered. The presence of small amount of MgO (Figs. 2,4) and the various types of defect features (Fig. 5) in present sample are perhaps responsible for the above observation. Mostly the *nano*-metric inclusions are not seen in *XRD* but can be visualized in *TEM* studies. Small *nano*-metric impurities or the ensuing disorder is though not favorable for better $\rho^{300K}$ and *RRR* values, the same might prove to be useful in pinning the vortices and hence improving the $J_c(H)$ of $MgB_2$. This we will discuss in conjunction with the magnetization results of the present study. The $\rho(T)$ plot of present $MgB_2$ is combination of the linear metallic part at higher *T* and the power law close to $T_c^{onset}$. For quantitative analysis the resistivity $\rho(T)$ can be defined as the sum of effective resistivity between the grains ($\rho^{gb}$) and the resistivity of grains ($\rho^{g}$) implying;

$$\rho(T) = \rho^{gb} + \rho^{g} = \rho^{gb} + \rho_0^{g} + \rho_{ph}^{g}(T) \qquad (1)$$

Here, $\rho_0^{g}$ is the residual resistivity and the $\rho_{ph}^{g}(T)$ is the term due to the scattering with the phonons, given by;

$$\rho_{ph}^{g}(T) = (m-1) \times \frac{d\rho}{dT} \times \Theta_D \left(\frac{T}{\Theta_D}\right)^m \times J_m\left(\frac{T}{\Theta_D}\right) \qquad (2)$$

$$\text{where, } J_m\left(\frac{T}{\Theta_D}\right) = \int_0^{\frac{\Theta_D}{T}} \frac{x^m\,dx}{(e^x - 1)(1 - e^{-x})} \qquad (3)$$

Here $\Theta_D$ is the Debye temperature and m = 3 to 5. In limiting case of $T < 0.1\Theta_D$ equation (1) becomes

$$\rho(T) = \rho_0 + AT^m \text{ (}A\text{ is the constant)} \qquad (4)$$

At $T = 0.1\Theta_D$, equation 4 deviates from the experimental $\rho(T)$ plots [33, 39, 41].

The experimental plot is fitted using power law, $\rho(T) = \rho_o + A\,T^m$, where $m = 3$ and $\rho_o$ is the residual impurity scattering part which is independent of temperature. The value of $\rho_o$ is taken from experimental plot as 33.65μΩ-cm. The fitted and observed experimental plots are depicted in Fig. 7. The fitted power law plot is found to deviate from the experimental data at around 140K. This yields $\Theta_D$ = 1400K, within the assumption that the power law fitting deviates at $T = 0.1\Theta_D$. It is worth mentioning here that we followed the same procedure for fitting as in refs. 38-40. The values of $\Theta_D$ = 1400 K and $\rho_o$ = 33.65μΩ-cm for present $MgB_2$, when compared with that in ref. 33 the former is 1050K and later 39.7μΩ-cm. The $\rho_o$ values are though comparable, the $\Theta_D$ is relatively larger in present case. However we would like to mention that present value of $\Theta_D$, is in good accord with the one obtained from thermoelectric power $S(T)$ fitting of our data and others [33,42-45], to be discussed in next section.

### 3D. Magneto transport

Resistivity versus temperature under magnetic field $\rho(T)H$ plots of present $MgB_2$ compound, are given in Fig. 8 and Fig. 9. The $\rho(T)H$ measurements show $T_c(\rho \rightarrow 0)$ at 37 and 15.4K respectively in zero and 80kOe respectively. The superconducting transition temperatures ($T_c$) are defined by that temperatures at which $\rho \rightarrow 0$. The $\rho(T)H$ plots are shown in Fig. 8, while Fig.9 depicts the $d\rho/dT(T)H$ plots in various applied field of up to 80kOe. Single $d\rho/dT(T)$ peaks are seen for all measurements under various applied fields. Single peak corresponds to one step transition from normal state to superconducting state and the broadening of the d$\rho$/d$T$ peaks after applying the higher magnetic field denotes toward the broadening of the transition width with increasing the magnetic field. Fig. 10 represents the $H_{c2}$ versus $T$ plot, where upper critical field ($H_{c2}$) is defined from the 90% of the resistive transitions [46] and is marked in Fig. 8. Though the present sample is neither doped with nano-particles nor synthesized by special techniques still its, $H_{c2}(T)$ plot is comparable with other reports [17,20-22]. We believe the presence of various defect structures and small un-reacted MgO in the compound to serve as effective pinning centers and hence improves the irreversibility line i.e., $H_{c2}(T)$ plot. It is again worth mentioning that though *XRD* showed only a minute presence of MgO (Fig.2) in studied sample, the detailed *TEM* analysis had shown (Figs. 4 and 5), various other types of defect features in the $MgB_2$ matrix. These defect features along with *nano*-MgO precipitates and 3-4*nm* thick amorphous layers between the grain boundaries of otherwise crystalline $MgB_2$ may be responsible for better performance, i.e., $H_{c2}$ = 80kOe at 21.5 K. For example in a recent report on *nano*-scale C doped $MgB_2$ tapes, $T_c$ ($\rho$=0) of 13K is observed at 130kOe [47]. Also in C doped [48] and neutron irradiated [49] $MgB_2$ bulk samples the $T_c(\rho \rightarrow 0)$ of around 17K is seen in 10 Tesla field. As far as $H_{c2}$ is concerned the same is around 10 Tesla at 21.5K [48]. In our case (pure $MgB_2$) and in ref. 48 ($MgB_{2-x}C_x$, x = 0.076) the same definition of $H_{c2}$ (90% of the resistive transitions) is used.

### 3E. Thermoelectric power Analysis

The $S(T)$ plot of present $MgB_2$ is shown in Fig. 11. The absolute value of $S$ is positive, which indicates towards the hole type conductivity in this system. The binding energy of Mg in $MgB_2$ is reported less than as expected for $Mg^{+2}$ [50]. Lowering of Mg charge results in $\sigma$ to $\pi$ electron transfer in Boron (B) giving rise to holes in the $\sigma$ band [50]. It is the $\sigma$ to $\pi$ electron transfer which gives rise to hole superconductivity in $MgB_2$. Superconducting transition ($T_c$) is seen as $S$=0 at 38K, corroborating the $R(T)$ data. For most of superconductors including $MgB_2$, small bump is reported just above the transition temperature. $S(T)$ has mainly the contributions from electrons and phonons. The room temperature thermoelectric power $S^{300K}$ is around 8μV/K, which is comparable to that as reported in refs. 33, 42, 44 and 45. However the shape of $S(T)$ plots is slightly different in all the references, in particular the change in slope of $S(T)$ before onset of $T_c$. Shape of our $S(T)$ plot is more close to that as in ref. 33, 43 and 45, but slightly different from that as in refs. 42 and 44.

The normal state of thermopower in $MgB_2$ in the range of ~40 to ~100K may be explained by

$$S = S_d + S_g = \frac{C_e}{ne} + \frac{C_{ph}}{3ne} = A.T + B.T^3 \qquad (5)$$

Where $S_d$ is the electronic diffusion term and $S_g$ is the phonon drag term. The coefficients A and B are given by $\gamma/ne$ and $\frac{K_B}{e}\frac{1}{n_a}\frac{4\pi^2}{5}\frac{1}{\Theta_D{}^3}$ respectively with the number of free electrons per atom $n_a$.

Generally the phonon relaxation time for interaction with other phonons and impurities is much larger than the relaxation time for the phonon-electron interaction. This is quite important below the Debye temperature ($\Theta_D$).

The theoretical curve of $S(T)$ is drawn using the equation $S = AT + BT^3$. By fitting the experimental curve with theoretical equation, we found the fitting parameters A = $1.921*10^{-2}$ $\mu V/K^2$, B = $0.936463*10^{-6}$ $\mu V/K^4$.

In our sample, the continuous line is best fitted in the range $3000K^2 < T^2 < 11000\ K^2$, for the plot (Fig.12) of $S/T$ as the function of $T^2$ i.e., from 55 to 105 K. This shows that in this region the $S_d$ and $S_g$ are dominant. Above 105K $S_d$ dominates. Below 55K, the linearity of $S/T$ vs $T^2$ is not observed and hence it is concluded that below this temperature both the $S_g$ and $S_d$ play a role. Correlation/fluctuation effects dominate as evident from increase in $S/T$. The phonon drag term $S_g$, is proportional to $C_{ph}$ as long as the electron – phonon interaction is the main scattering mechanism for phonons. As temperature increases, the phonon-impurity and phonon-phonon process becomes important. In this situation phonons do not transfer momentum only to electrons and the phonon drag fails.

In reported data and analysis of $S(T)$ of $MgB_2$, we found that though high temperature (> 100K) behavior is same in most of them [33, 42-45], the usual low $T$ (< 100K) near constant $S(T)$ part is not seen is some of them [43, 44], which is true in some more recent reports as well [51,52]. The $S(T)$ of the single crystals of $MgB_2$ also shows nearly the similar behavior to that as in bulk at least in *ab*-direction [53,54].

As far as the upper temperature limit of fitting is concerned the $S(T)$ in principle can be fitted in the range of $T_c < T < 0.1\Theta_D$. The estimated Debye temperature ($\Theta_D$), is around 1412 K, hence fitting up to 105K is quite reasonable. Besides the $\Theta_D$, the values of Fermi energy ($\varepsilon_F$) and carrier density ($n$) are also estimated from the $S(T)$ equations and A/B constants. We have compared our values of the fundamental parameters of present $MgB_2$ with reported data in ref. 33, 42 and 43 (Table 1). The estimated Debye temperature ($\Theta_D$), and carrier density ($n$) are found to be comparable ($\Theta_D$ = 1412K and $n$ = $3.81*10^{28}/m^3$).

### 3F. Thermal conductivity $\kappa(T)$ behavior

Figure 13 shows the thermal conductivity (κ) data from 300 K down to low temperatures. The absolute values obtained are nearly half or even one third to that as reported in references 33, 42, 45, 51 and 52. Interestingly the $\kappa(T)$ plot shape is very different in both these references at least in low $T$ regime of below 100K. In ref. 33, 42, 45, 51 and 52, the $\kappa(T)$ is shown linear above 100K with a dip in later at close to room temperature. At low $T$ say below 50K, the $\kappa(T)$ goes sharply up in ref. 42 before decreasing with a peak type shape. This low $T$ behavior is completely missed in ref. 33 where $\kappa(T)$ is seen linear above say 100K. Further no sign of $T_c$ is seen in $\kappa(T)$ plots shown in references 33, 38, 42, 45, 51 or 52. Also the absolute value of κ remains though > 10 W/mK in reference 42 the same is much less in reference 33 and 38 at all studied temperatures. Hence we can say that not only the shape of the $\kappa(T)$ plots but the absolute values are also very different from each other in reported literature.

Now we discuss $\kappa(T)$ behavior of present $MgB_2$ sample which is shown in Fig. 13. $\kappa(T)$ exhibits a hump like structure with κ increasing down to nearly 150K. Though similar small hump is seen in reference 42 close to room temperature, in our case the same is very broad and extended down to 150K. Our $\kappa(T)$ seemingly follows dominantly the increasing electronic part of κ down to 150K with $\kappa_{el} \sim T^{-m}$ with m ~ 2, and the phonon contribution ($\kappa_{ph}$) seems to be marginalized. Below around 150K, the $\kappa(T)$ decreases linearly with $T$, which is similar to that as in refs 33, 38 and 45 but altogether different than that as in ref 38, 51 and 52. An interesting fact is that in our $\kappa(T)$ plot clear sign of $T_c$ is seen. For more clarity the same is shown as extended $\kappa(T)$ plot in inset of Fig. 13. Although the 'peak effect' below $T_c$ is marginally present, the curves drawn to guide the eye clearly distinguish the temperature behaviors on the two sides of $T_c$. In earlier reports the κ($T$) of $MgB_2$ did not exhibit any sign of superconducting transition [13, 33, 38, 42, 45, 51, 52, 55, 56]. However on the contrary, recently the κ($T$) data on pure and substituted $MgB_2$ single crystals had shown an anomaly in thermal conductivity below $T_c$ [57]. Our present results are in confirmation with the κ($T$) anomaly at $T_c$ in ref. 57.

Because of the differences in shape and absolute value of the present κ($T$) in comparison to and also among the earlier reports [13, 33, 38, 42, 45, 51, 52, 55, 56], we have not attempted to fit our data to the known individual contributions. Here again we would like to mention that these variations in the absolute κ($T$) values are simply due to the poor grains connectivity in various samples, see ref. 29. This issue is discussed previously in resistivity results section 3.C. The intriguing point is that though the $\rho^{300K}$ and $S^{300K}$ values are comparable to each other in present study and ref. 33, 34, 45 and 56 for normal route synthesized $MgB_2$ samples, the values of κ($T$) are different, and not exactly scaling with the disorder/porosity. For example $\rho^{300K}$ - $\rho^{40K}$ is ~ 44μΩ.cm (present), 35μΩ.cm (ref56), and 70μΩ.cm (ref.33), on the other hand the $\kappa^{300K}$ values are 3.6 W/mK(present), 15 W/mK(ref56), and 9W/mK(ref.33); clearly not scaling with the extent of disorder. This is unlike the $\rho^{300K}$ and $S^{300K}$ values. This may be due to different impacts of porosity of the sample on thermal and electrical conduction [58]. In such situations though $\rho^{300K}$ and $S^{300K}$ values could scale with extent of disorder but the $\kappa^{300K}$ may not. Seemingly our present sample has more porous regions than in ref. 34 and 56, and hence its $\kappa^{300K}$ value is smaller than that as in ref. 33,34 and 56, despite being having similar extent of disorder. Interestingly though the pronounced maximum indicates towards very clean sample, its absolute value of κ($T$) is the smallest. This needs further examination and as such the explanation given above in regards to different impact of porosity on thermal and electrical conduction may not hold.

In another report the grains connectivity is controlled by increasing the density of polycrystalline $MgB_2$ from 1.8 gram/cm$^3$ to 2.4 gram/cm$^3$. Thus achieved dense samples showed residual resistivity to be around 0.5μΩ.cm and thermal conductivity to be as high as 215 W mK$^{-1}$[34]. Note that in our sample we get residual resistivity to be 33.65μΩ.cm, see section 3.C. Also the authors from ref. 34 observed earlier the same to be 39.7 μΩ.cm for a normal route synthesized sample [33]. The highly dense (90% of theoretical density) samples used in ref. 34 were synthesized by special synthesis route [59].

**3G. Magnetization studies**

Figure14 depicts the dc susceptibility versus temperature $\chi(T)$ plots in an applied field of 10 Oe, in both zero-field-cooled (*ZFC*) and field-cooled (*FC*) situations. It is evident from this figure that $MgB_2$ undergoes a sharp superconducting transition (diamagnetic, $T_c^{dia}$) at 37.2K within less than 1K temperature interval, without any unusual rounding occurring down to 5K. In fact, the diamagnetic signal remains more or less constant below 36K down to 5 K. It is found that the superconducting citical temperature ($T_c^{dia}$) being seen from $\chi^{ZFC}(T)$ measurements is in agreement with the $T_c$ as seen from $d\rho/dT(T)$ peaks. The $\chi^{FC}(T)$ plot is rather merged with the zero base line. In fact the $\chi^{FC}(T)$ is very small and close to the instrument detection limit of $10^{-6}$ emu. We found that this has happened in most of pinned $MgB_2$ compounds with the *FC* signal being very small in comparison to the *ZFC*. Seemingly the present $MgB_2$ itself is in the pinned state. Interestingly various defect structures and *nano*-metric impurities (Figs. 2-5) might act as effective pinning centers for flux vortices in the present sample. Very low value of $\chi^{FC}$ observed in comparison to $\chi^{ZFC}$ signal (Fig.14) in superconducting state i.e., below say 38 K is an indication in this direction. As far as superconducting volume fraction is concerned, in the case of strong pinning, the $\chi^{FC}$ is too small and hence its estimation is near impossible. One may, however, estimate the volume fraction from $\chi^{ZFC}$, but that would clearly be incorrect, as this would include a large contribution arising from shielding currents. Besides the *nano*-metric MgO and various defects, the 3-4*nm* thick amorphous layers (Fig.5) between the grain boundaries of $MgB_2$ might also be responsible for the pinning in presently studied compound.

Figure 15 shows the high field magnetization results for present $MgB_2$ compound at 4, 10, and 20K with applied fields ($H$) of up to 80kOe. Some fluxoid jumps are seen at $T$ = 4K below 10kOe field. Very recently [60] we found that these fluxoid jumps are intrinsic in nature and occur in high $J_c$ $MgB_2$ samples. The magnetization $M(H)$ grows (as usual) slowly with $H$ and falls sharply to near zero moment value, and further grows again in a common way. The flux avalanches were seen quite symmetric in both increasing/decreasing the field in all four quadrants of the $M(H)$ loops [60]. The dynamics of these flux avalanches is discussed elsewhere [60-62], and as such need not to be elaborated here. Fundamentally very low heat capacity and relatively high $J_c$ values in $MgB_2$ are seemingly the cause for the observed complex vortex dynamics [60-62].

We estimated the $J_c$ of our sample from data in Fig. 15, by invoking the Bean's critical state model, and the results are plotted in Fig. 16. The near cylindrical shape sample used in the study has the dimensions 0.742 cm x 0.340 cm x 0.162 cm. We used the formula

$$J_c = 20 \times \Delta M \ / \ a(1-a/3b) \qquad \text{where } a < b \qquad [6]$$

Here $\Delta M = |M_+| - |M_-|$ which comes from the $M(H)$ loops and a is the thickness & b is the width of the bar shaped sample. In this manner the $J_c(H)$ values are estimated and are plotted in main panel of Fig. 16.

We got an estimate of $J_c$ of up to around 1.05 x $10^5$A/cm$^2$ at 4 and 10K in 10kOe field. The $J_c(H)$ below 10kOe is not plotted, because of flux avalanches observed in that region, mentioned above. It is important to note that while comparing the critical currents of different superconducting materials, the size and shape of the measured sample may influences the outcome [17, 64-66]; see interesting comparisons and suggestions in ref. 67. With an increase in field the $J_c(H)$ comes down as expected for any other superconductor. The $J_c(H)$ comes down to $10^2$A/cm$^2$ at 42, 68 and 78kOe at 20, 10 and 4 K respectively. In fact these values of applied fields can be related to the irreversibility field ($H_{irr}$), i.e. the field at which a superconductor permits most of the external field to pass through it, or other wise expelled in reverse direction. To estimate the $H_{irr}$ we show the extended $M(H)$ plots in inset of Fig. 16. It is observed that $H_{irr}$ for the studied sample is around 78, 68 and 42kOe at 4, 10 and 20K respectively.

As far as the comparison of the $J_c$ value ($10^5$ A/cm$^2$) is concerned with literature, the same seems to be competitive without any special heat treatments or the external doping in $MgB_2$. The studied sample of $MgB_2$ though contains internal flux pinning centers in the form of *nano* defect structures, *n*-MgO and the *nano*-metric grain boundary precipitates but is not doped extrinsically with *nano* sized additives viz. *n*-SiC [20, 21], *n*-Diamond [23], or *n*-carbon tubes [22] etc. The internal pinning centers in the form of *nano*-defects, *n*-MgO and *nano*-metric grain boundary precipitates etc. had earlier not been studied, which seemingly contribute to $J_c(H)$ performance in $MgB_2$ superconductor. Worth mentioning is the fact that critical current density $J_c(H)$ of *nano*-particle doped $MgB_2$ had already reached exceeding $10^4$ A/cm$^2$ in applied fields of up to 100-140kOe at 4.2 K [47,67,68]. These samples of $MgB_2$ contained apparently both intrinsic and extrinsic pinning centers and hence exhibited dramatic $J_c(H)$ performance.

In order to improve the superconducting performance of our pure $MgB_2$, very recently we doped *nano*-SiC & *nano*-Diamond particles in Pure $MgB_2$ [71,72]. The results of best performance [71,72] optimally doped (5wt% *n*-Sic and 3wt% *n*-Diamond) samples are summarized here for irreversibility field ($H_{irr}$), critical current density $J_c(H)$ & flux pinning and compared with the similarly synthesized pristine $MgB_2$, which is being discussed in previous sections. The *nano*-SiC and *nano*-Diamond samples are synthesized at the same optimized temperature of 850 $^0$C in Argon atmosphere as being used for pure $MgB_2$. The superconducting performance for both *nano*-SiC and *nano*-Diamond added $MgB_2$ is checked, and the results are presented in Figs. 17 and 18. To see the doping effect of these on superconducting performance the $M(H)$ & $H_{irr}(T)$ plots at 10K are shown in fig. 17. The main panel of fig.17 clearly demonstrates that the $M(H)$ loop for pristine sample is almost closed at 68 kOe, whereas the same is opened up to 120 kOe and 98 kOe for *n*-D3 and *n*-SiC5 doped samples respectively. This is the situation at 10K, the $H_{irr}$ value is about 140 kOe and 113 kOe for the *n*-D3 and *n*-SiC5 doped samples respectively at 5K, see inset of fig. 17. The critical current density $J_c(H)$ is calculated and plotted in the inset of Fig. 18 along with flux pinning plots in main panel of the same at 10K. At low fields the $J_c$ values are nearly $10^5$ A/cm$^2$ for all the three samples. As we see form 18 the value of $J_c$ at 10K is clearly higher by about an order of magnitude or so than pristine $MgB_2$ for both *n*-SiC5 and *n*-D3 doped samples at 60 kOe applied fields. This is because of role of *nano* dopants as the effective pinning centers in host sample and hence improving the superconducting performance at higher fields. Also the possible substitution of Carbon at Boron site, being available from some broken *n*-SiC and *n*-D $MgB_2$ could have played an important role [71,72]. In fact in recent years, *n*-SiC/*n*-diamond/*n*-carbon-tubes/*n*-carbon doped $MgB_2$ superconductor had yielded high dividends [12-23, 67-70]. It is known that substitution of C at B-site in $MgB_2$ creates disorder in Boron plane and thus improves the superconducting performance [67-70]. This when clubbed with effective available *nano*-pinning centers, gives rise to high superconducting performance in applied fields [67-72]. For more detailed analysis see ref. 71 for $MgB_2$ + *n*-SiC and ref. 72 for $MgB_2$ + *n*-Diamond.

Finally worth mentioning is the fact that very recent results on dense wires and tapes of *nano*-particle added $MgB_2$ had further improved the superconducting performance of this compound [73-75]. It seems that the porous nature of various polycrystalline $MgB_2$ samples has crucial role in its various physical properties, including superconducting parameters, due to its poor grains connectivity [29]. This need to be tackled intelligently through various synthesis steps, such as two step heat treatments, aligned films or compact powder in tube tapes and wires etc. [73].

## SUMMARY AND CONCLUSIONS

We studied in detail various physical properties of bulk polycrystalline $MgB_2$ superconductor. In particular its, micro-structure, Raman spectroscopy, resistivity under magnetic field [$\rho(T)H$], thermoelectric power $S(T)$, thermal conductivity $\kappa$ $(T)$ and magnetization are studied in detail. Micro-structure of the compound showed the presence of various types of defect features with the inclusions of nearly spherical size (4-15*nm*) MgO in matrix of $MgB_2$. Other micro-structural features, which were frequently seen during *TEM* analysis, are the presence of 3-4nm thick amorphous layers between the grain boundaries of otherwise crystalline $MgB_2$. The $\rho(T)H$ measurements confirmed $T_c(\rho=0)$ at 37K in zero field and 15.4K at 80kOe. The power law fitting of $\rho(T)$ gave rise to Debye temperature ($\Theta_D$) = 1400K. Theoretical fitting of $S(T)$ exhibited $\Theta_D$ = 1410K and carrier density of $3.81*10^{28}/m^3$. Thermal conductivity $\kappa(T)$ behavior of our sample is though very similar to the literature reports but much lower in magnitude. Also our $\kappa(T)$ clearly exhibited a jump at 38K, i.e., at $T_c$, which was missing in some earlier reports. Critical current density ($J_c$) of up to $10^5 A/cm^2$ in 1-2T (Tesla) fields at temperatures ($T$) of up to 10K is seen from magnetization measurements invoking the Bean's critical state model. For technical applicability, *nano*-additives like *nano*-SiC and *nano*-Diamonds can be added to the title compound and its performance can be enhanced dramatically [71,72].

**ACKNOWLEDGEMENT**

Authors thank Prof. J.M. Rowell for his suggestions and various discussions. Mr. A.K. Sood from *SEM* Division of *NPL* is acknowledged for providing us with the *SEM* micrographs. Dr. Rajeev Ranjan from *BHU* is acknowledged for helping us in preliminary Reitveld fitting of our pure sample. Dr. Rajeev Rawat and Dr. Vasant Sathe from *CSR*-Indore are acknowledged respectively for the resistivity under magnetic field and Raman measurements. Mr. Kranti Kumar and Dr. A. Banerjee are acknowledged for the high field magnetization measurements. Further *DST*, Government of India is acknowledged for funding the 14 Tesla-PPMS-VSM at CSR, Indore. Arpita Vajpayee and Monika Mudgel would like to thank the *CSIR* for the award of Junior Research Fellowship to pursue their *Ph. D* degree Authors from *NPL* appreciate the interest and advice of Prof. Vikram Kumar (Director) in the present work.

**REFERENCES**

1. J. Nagamatsu, N. Nakagawa, T. Muranaka, Y. Zenitani, and J. Akimitsu, Nature (London) **410** (2001) 63.

2. Cristina Buzea and Tsutomu Yamashita, Sup. Sci. and Tech. **14** (2001) R 115.

3. T. Muranaka, Y. Zenitani, J. Shimoyama and J. Akimitsu, Frontiers in superconducting materials, Ed. By A.V. Narlikar, Springer-Verlag Germany, p.937 (2005).

4. Thomas Dahm, Frontiers in superconducting materials, Ed. By A.V. Narlikar, Springer-Verlag Germany, p.983 (2005).

5. S.L. Bud'ko, G. Lapertot, C. Petrovic, C.E. Cunningham, N. Anderson and P.C. Canfield, Phys. Rev. Lett. **86** (2001) 1877.

6. Hinks D G, Claus H and Jorgensen J D, Nature **411** (2001) 457.

7. A.Y. Liu, I.I. Mazin and J. Kortus, Phys. Rev. Lett. **87** (2001) 087005.

8. H.J. Choi, D. Roundy, H. Sun, M.L. Cohen and S.G. Louie, Phys, Rev. B **66** (2002) 020513.

9. M. Putti, R. Vaglio and J.M. Rowell, Sup. Sci. and Tech. **21** (2008) 043001.

10. O.F. de Lima, R.A. Ribeiro, M.A. Avila, C.A. Cardoso, and A.A. Coelho, Phys. Rev. Lett. **86** (2001) 5974.

11. S.B. Samanta, H. Narayan, A. Gupta, A.V. Narlikar, T. Muranaka, and J. Akimitsu, Phys. Rev. B. **65** (2002) 92510.

12. J. Kortus, I.I. Mazin, K.D. Belaschenko, V.P. Antropov, and L.L. Boyer, Phys. Rev. Lett. **86** (2001) 4656.

13. E. Bauer, Ch Paul, St Berger, S Majumdar, H Michor, M. Giovannini, A. Saccone and A Bianconi, J. phys: Condens. Matter **13** (2001) L487.

14. A.V. Sologubenko, J. Jun, S.M. Kazakov, J. Karpiniski, and H.R. Ott, Phys. Rev. B. **65** (2002) 18505R.

15. Ch. Walti, E. Felder, C. Dengen, G. Wigger, R. Monnier , B. Delley and H.R. Ott, Phys. Rev. B **64** (2001) 172515.

16. F. Bouquet, R.A. Fisher, N. E. Phillips, D. G. Hinks and J.D. Jorgensen Phys. Rev. Lett. **87** (2001) 047001.

17. A. Matsumoto, H. Kumakura, H. Kitaguchi, and H. Hatakeyama, Sup. Sci. and Tech. **16** (2003) 926.

18. G. Profeta, A. Continenza, A. Floris and S. Massidda, Phys. Rev. B **67** (2003) 174510.

19. S.X. Dou, S. Soltanian, W.K. Yeoh and Y. Zhang, IEEE Trans. on applied supercon. **15** (2005) 3219.

20. S.X. Dou, S. Soltanian, X.L. Wang, P. Munroe, S.H. Zhou, M. Ionescu, H.K. Liu, and M. Tomsic, Appl. Phys. Lett. **81** (2002) 3419.

21. S.X. Dou, V. Braccini, S. Soltanian, R. Klie, Y. Zhou, S. Li, X.L. Wang, and D. Larbalestier, J. Appl. Phys. **96** (2004) 7549.

22. W.K. Yeoh, J.H. Kim, J. Horvat, S.X. Dou and P. Munroe, Sup. Sci. and Tech. **19** (2006) L5.

23. C.H. Cheng, H. Zhang, Y. Zhao, Y. Feng, X.F. Rui, P. Munroe, H. M. Zheng, N. Koshizuka, and M. Murakami, Sup. Sci. and Tech. **16** (2003) 1182.

24. C.M. Franco, B. Ferreira, C.A.M. Santos, L. Ghiveldar, H.J.I. Filho, and A.J.S. Machado, Physica C **408**, (2004) 130.

25. N.N. Kolesnikov and M.P. Kulakov, Physica C **363**, (2002)166.

26. K.P. Singh,V.P.S. Awana, Md. Shahabuddin, M. Husain, R.B. Saxena, Rashmi Nigam, M.A. Ansari, Anurag Gupta, Himanshu Narayan, S.K. Halder, and H. Kishan Mod. Phys. Lett. B. **27**, (2006) 1763.

27. V.P.S. Awana, Rajeev Rawat, Anurag Gupta, M. Isobe, K.P. Singh, Arpita Vajpayee, H. Kishan, E. Takayama-Muromachi and A.V. Narlikar, Solid State Communication **139** (2006) 306.

28. K.P.Singh, V.P.S. Awana, Md. Shahbuddin, Intikhab A Ansari, M. Husain, and H. Kishan, To appear in Cryogenics (2007).

29. John M Rowell, Sup. Sci. and Tech. **16** (2003) R17-R27.

30. Rodrigues-Carvajal J "FULLPROF" 2000. A Rietveld Refinement and Pattern Matching Analysis Program. Laboratoire Leon Brillouin (CEA-CNRS) France.

31. Rajeev Ranjan, V. P. S. Awana, H. Kishan and Dhananjay Pandey*, Private communication*.

32. Y. Zhu, L. Wu, V. Volkov, Q. Li, G. Gu, A. R. Moodenbaugh, M. Malac, M. Suenaga, and J. Tranquada, Physica C **356** (2001) 239.

33. M. Putti, E. Galleani d'Agliano, D. Marre, F. Napoli, M. Tassisto, P. Manfrinetti, and A. Palenzona, Studies of High Temperature Superconductors (Ed.A.Narlikar), Nova Science Publishers, NY, Vol.**38** (2002) 313; also at Euro Phys. J. B **25** (2002) 439.

34. M. Putti, V. Braccini, E. Galleani, F. Napoli, I. Pallecchi, A.S. Siri, P. Manfrinetti, and A. Palenzona, Supercond. Sci. and Tech. **16** (2003) 188.

35. T. Matsui, S. Lee and S. Tajima, Phys. Rev. B. **70** (2004) 024504.

36. B. Renker, K.B. Bohnen, R. Heid, D. Ernst, H. Schober, M. Koza, P. Adelmann, P. Schweiss, and T. Wolf. Phys. Rev. Lett. **96** (2006) 077003.

37. T. Matsui, Physica C **456** (2007) 102.

38. M. Putti, V. Braccini, E. Galleani d' Agliano, F. Napoli, I. Pallechi, A.S. Siri, P. Manfrinetti and A. Palenzona, Phys. Rev. B **67** (2003) 064505.

39. A. Bharathi, S.J. Balaselvi, S. Kalavathi, G.L.N. Reddy, V.S. Sastry, Y. Hariharan, and T.S. Radhakrishnan, Physica C **370** (2002) 211.

40. P.C. Canfield, D.K. Finnemore, S.L. Bud'ko, J.F. Ostenson, G. Lapertot, C.E. Cunningham, and C. Petrovic, Phys. Rev. Lett. **86** (2001) 2423.

41. J.M. Ziman, Electronics and Phonons, Clarendon Press, Oxford (1960); R. D. Barnard, Thermoelectricity in metals and alloys, Taylor & Francis Ltd., London (1972).

42. J. Mucha, M. Peckela, J. Szydlowska, W. Gadomski, J. Akimitsu, J-F. Fagnard, P. Vanderbemden, R. Cloots and M. Ausloos, Sup. Sci. Tech. **16** (2003) 1167.

43. B. Lorenz, R.L. Meng, Y.Y. Xue, and C.W. Chu, Phys. Rev. B. **64** (2001) 052513.

44. J.S. Ahn, E.S. Choi, W. Kang, D.J. Singh, M. Han and E.J. Choi, Phys. Rev. B **65** (2002) 214534.

45. Bhaskar Gahtori, Ratan Lal, S.K. Agarwal, Y.K. Kuo, K.M. Shivkumar, J.K. Hsu, J.Y. Lin, Ashok Rao, S.K. Chen and J.L. MacManus-Driscoll, Phys. Rev. B **75** (2007) 184513.

46. A. Gurevich, S. Patnaik, V. Braccini, K.H. Kim, C. Mielke, XSong, L.D. Cooley, S.D. Bu, D.M. Kim, J.H. Choi, L.J. Belenky, J. Giencke, M.K. Lee, W. Tian, X.Q. Pan, A. Siri, E.E. Hellstrom, C.B. Eom and D.C. Larbalestier, Supercond. Sci. and Tech. **17** (2004) 278.

47. Y Ma, X. Zhang, S. Awaji, D. Wang, Z. Gao, G. Nishijima and K. Watanabe, Supercond. Sci. and Tech. **20** (2007) L5.

48. R.H.T. Wilke, S. L. Bud'ko, P.C. Canfield and D.K. Finnemore, Phys. Rev. Lett. **92** (2004) 217003.

49. M. Putti, V. Braccini, C. Ferdeghini, F. Gatti, G. Grasso, P. Manfrinetti, D. Marre. A. Palenzona, I. Pallacchi, C. Tarantini, I. Sheikin, H.U. Aebersold and E. Lehmann, Appl. Phys. Lett. **86** (2005) 112503.

50. A. Talpatra, S.K. Bandyopadhyay, Pintu Sen, P. Barat, S. Mukherjee, and M. Mukherjee, Physica C **419** (2005) 141.

51. M.E. Yakinci, Y. Balci and M.A. Aksan, Physica C **408-410** (2004) 684.

52. M.A. Aksan, M.A. Yakinci and A. Güldste, J. Alloys and compounds **424** (2006) 33.

53. T. Masui, K. Yoshida, S. Lee, A. Yamamoto and S. Tajima1, Phys. Rev. B **65** (2002) 214513.

54. T. Plackowski, C. Sułkowski, J. Karpinski, J. Jun and S.M. Kazakov, Phys. Rev. B **69** (2004) 104528.

55. T. Muranaka, J. Akimitsu and M. Sera, Phys. Rev. B **64** (2001) 020505.

56. A.V. Sologubenko, J. Jun, S.M. Kazakov, J. Karpiniski, and H.R. Ott, Phys. Rev. B. **66** (2002) 014504.

57. A.V. Sologubenko, N.D. Zhidaglo, S.M. Kazakov, J. Karpiniski, and H.R. Ott, Phys. Rev. B. **71** (2005) 020501R.

58. J.M. Rowell *Private communication* (2008).

59. A. Palenzona, P. Manfrinetti, and V. Braccini, INFM Patent no. TO2001A001098.

60. I. Felner, V.P.S. Awana, Monika Mudugal and H. Kishan, J. Appl. Phys. **101** (2007) 09G101.

61. J. Gaojie, J. Xu, J.C. Grivel, A.B. Abrahmsen, and N.H. Andersen**,** Physica C **406** (2004) 95.

62. Victor Chabanenko, Roman Puzniak, Adam Nabialek, Sergei Vasiliev, Vladimir Rusakov, Loh Huanqian, Ritta Szymczak, Henryk Szymczak, Jan Juan, Jansuz Karpiniski, and Vitaly Finkel, J. Low Temp. Phys. **130** (2003) 175.

63. V.P.S. Awana, M. Isobe, K.P. Singh, Md. Shahabuddin, H. Kishan, and E. Takayama-Muromachi Sup. Sci. and Tech**.** **19** (2006) 551.

64. S.X. Dou, S. Soltanian, Y. Zhao, E.Getin, Z. Chen, O. Scherbakova and J. Horvat, Sup. Sci. and Tech. **18** (2005) 710.

65. M. Delfany, X.L. wang, S. Slotanian, J. Horvat, H.K. Liu and S.X. Dou, Ceramics International **30** (2004) 1581.

66. J. Horvat, S. Soltanian, X.L. Wang and S.X. Dou, Appl. Phys. Lett. **84** (2004) 3109.

67. S.X. Dou, W.K. Yeoh, O. Shcherbakova, D. Wexler, Y. Li, Z.M. Ren, P. Munroe, S.K. Tan, B.A. Glowacki, and J.L. MacManus-Driscoll, Adv. Mater. **18** (2006) 785.

68. M. Eisterer, R. Muller, R. Schoppl, H.W. Webber, S. Soltanian and S.X. Dou, Supercond. Sci. & Tech. **20** (2007) 117.

69. A. Matsumoto, H. Kumakura, H. Kitaguchi, B.J> Senkowicz, M.C.Jewell, E.E. Hellstrom, Y. Zhu, P.M. Voyles and D.C. Larbalestier, Appl. Phys. Lett. **89** (2006) 132508.

70. A. Serquis, G. Serrano, S.M. Moreno, L. Civale, B. Maiorov, F. Balakirev and M. Jaime, Supercond. Sci. & Tech. **20** (2007) L12.

71. Arpita Vajpayee, V.P.S. Awana, G.L. Bhalla and H. Kishan, Nanotechnology **19** (2008) 125708.

72. Arpita Vajpayee, V.P.S. Awana, H. Kishan, A.V. Narlikar, G.L. Bhalla and X. L. Wang, J. App. Phy. **103** (2008) 07C708.

73. W.K. Yeoh ans S.X. Dou, Physica C **456** (2007) 170.

74. C.H. Cheng, Y. Yang, P. Munroe and Y. Zhao, Supercond. Sci. & Tech. **20**, (2007) 296.

75. S.C. Yan, G. Yan, L. Zhou, Y. Jia, H.H. Wen, and Y.F. Lu Supercond. Sci. & Tech. **20**, (2007) 377.

**FIGURE CAPTIONS:**

**Fig.1** X-ray diffraction pattern for various Argon annealed $MgB_2$ compound.

**Fig.2** Reitveld fitted and observed XRD patterns for 850 $^0$C Argon annealed $MgB_2$ compound.

**Fig.3 (a)** and **b)** Scanning electron microscope (SEM) pictures of 850 $^0$C Argon annealed $MgB_2$ compound.

**Fig.4(a**) Bright-field *TEM* micrograph of 850 $^0$C Argon annealed $MgB_2$, showing the presence of partial dislocations

**Fig.4(b)** Dark-field *TEM* micrograph of 850 $^0$C Argon annealed $MgB_2$, showing the presence of partial dislocations.

**Fig.4(c)** *TEM* micrograpgh of 850 $^0$C Argon annealed $MgB_2$, showing the presence of MgO spherical nano-particles of 4-15*nm*

**Fig**.**5(a)** and **(b)**: *TEM* details of 850 $^0$C Argon annealed $MgB_2$ gain boundary pictures showing the precipitates

**Fig. 6** Raman spectra for 850 $^0$C Argon annealed $MgB_2$ and $Mg_{0.6}Al_{0.4}B_2$ at room temperature.

**Fig.7** Experimental and fitted $\rho$ ($T$) plots for 850 $^0$C Argon annealed $MgB_2$

**Fig.8** $\rho$ ($T$) plots for 850 $^0$C Argon annealed $MgB_2$ with $H$ = 0 to 80kOe

**Fig. 9** d$\rho$/d$T$ ($T$) plots of 850 $^0$C Argon annealed $MgB_2$ at applying different fields range from 0 to 80kOe

**Fig. 10** Upper critical field as a function of the temperature as determined from the 90% of the resistive transitions $R(T)H$ for 850 $^0$C Argon annealed $MgB_2$

**Fig.11** Experimental $S(T)$ plot of 850 $^0$C Argon annealed $MgB_2$ fitted with diffusive and phonon terms

**Fig.12** Continuous linear fitted plot for $S/T$ vs. $T^2$ plot for 850 $^0$C Argon annealed $MgB_2$

**Fig.13** $\kappa(T)$ plot of 850 $^0$C Argon annealed polycrystalline $MgB_2$ pellet with superconducting transition at ~38K and a broad maximum around ~150K. Inset highlights the low temperature anomaly at $T_c$ and the linear $T$-dependence in the normal state.

**Fig.14** $\chi(T)$ plot of 850 $^0$C Argon annealed $MgB_2$

**Fig.15** $M(H)$ plots for 850 $^0$C Argon annealed $MgB_2$ sample at 4, 10 and 20K.

**Fig.16** $J_c(H)$ plot for 850 $^0$C Argon annealed $MgB_2$, the inset shows the extended $M(H)$ plots to mark the $H_{irr}$ of the sample

**Fig. 17** The $M(H)$ & $H_{irr}(T)$ plots at 10 K for 850 $^0$C Argon annealed pure, 5 wt% *n*-SiC and 3 wt% *n*-D added $MgB_2$ samples

**Fig. 18** The critical current density $J_c(H)$ for 850 $^0$C Argon annealed pure, 5 wt% *n*-SiC and 3 wt% *n*-D added $MgB_2$ samples at 10 K

**TABLE-1** $S(T)$ fitting parameters for present $MgB_2$ and their comparison with reported literature.

| Systems | A($\mu V/K^2$) | B($\mu V/K^4$) | $\varepsilon_F$ (ev) | $n_e$ ($10^{28}$)/$m^3$ | $\Theta_D$ (K) |
|---|---|---|---|---|---|
| Present $MgB_2$ | $1.921 \times 10^{-2}$ | $0.937 \times 10^{-6}$ | 1.91 | 3.81 | 1412 |
| Ref.33 | 2.0 x $10^{-2}$ | 1.30 x $10^{-6}$ | 1.82 | 5±2 | 1430 |
| Ref.43 | 1.5x$10^{-2}$ | 1.20x$10^{-6}$ | 2.50 | - | 1450 |
| Ref. 42 | $1.76 \times 10^{-2}$ | $1.26 \times 10^{-6}$ | 2.08 | 6 ± 2 | 1430 |

Figure 1

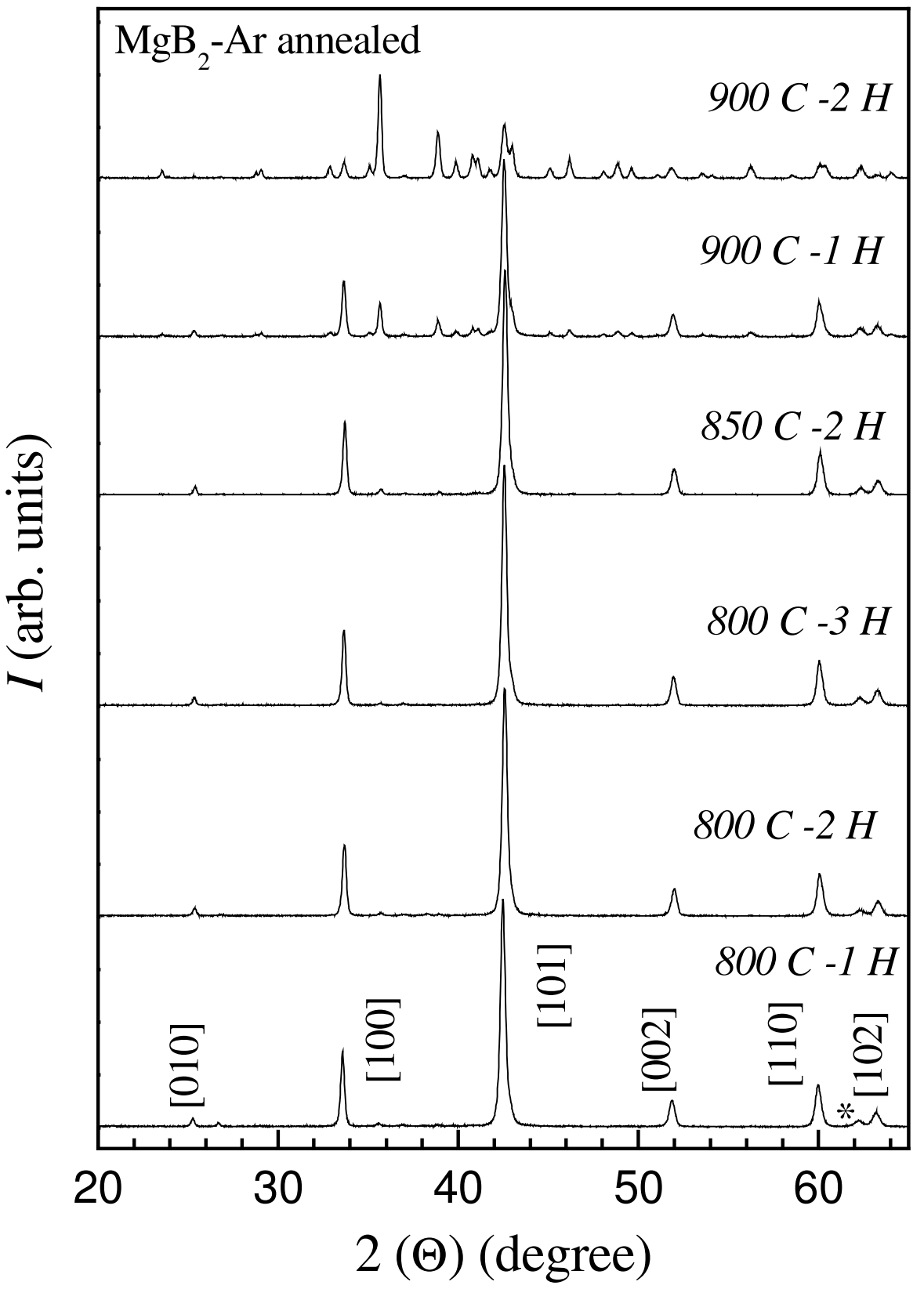

Figure 2

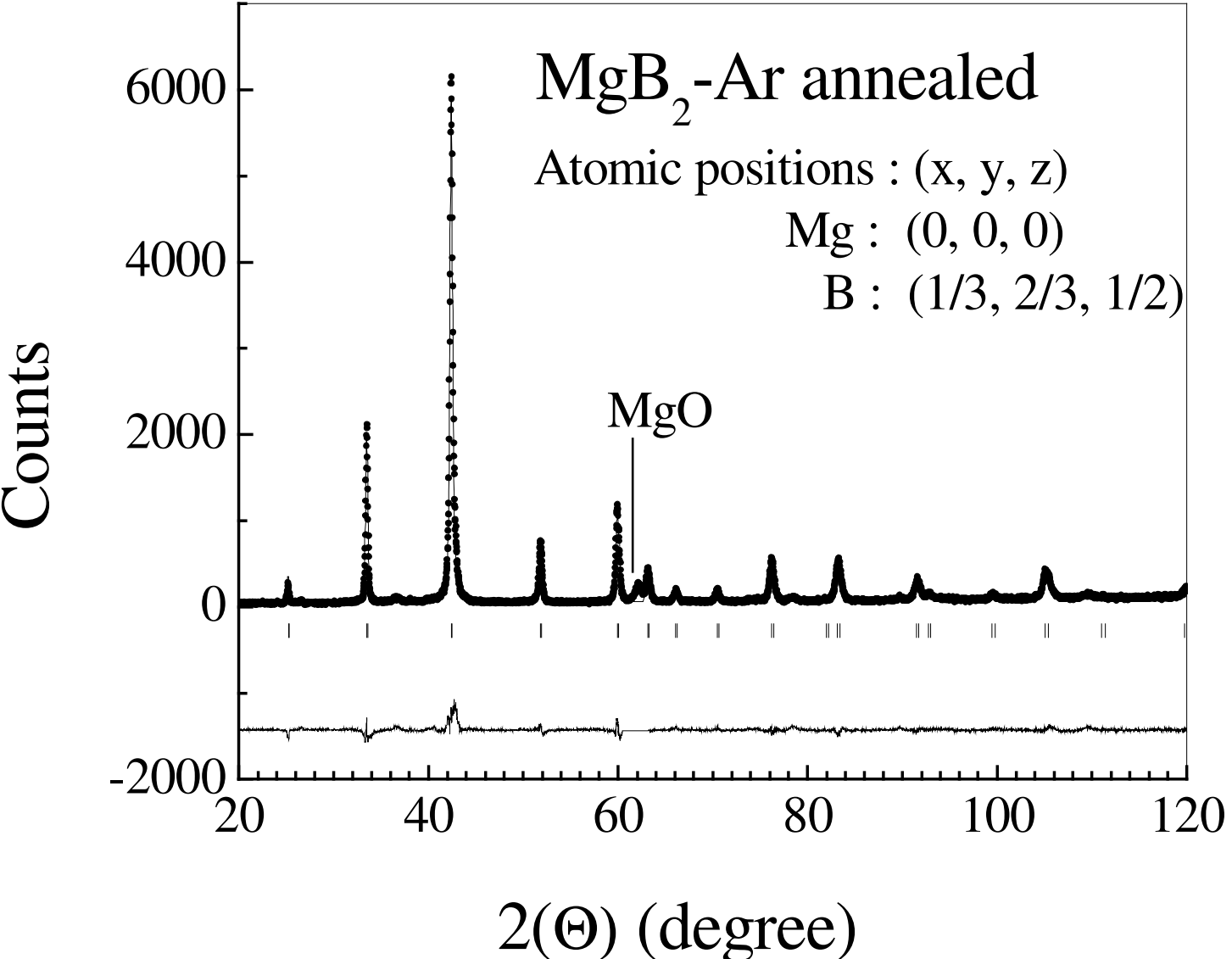

Figure 3 (a)

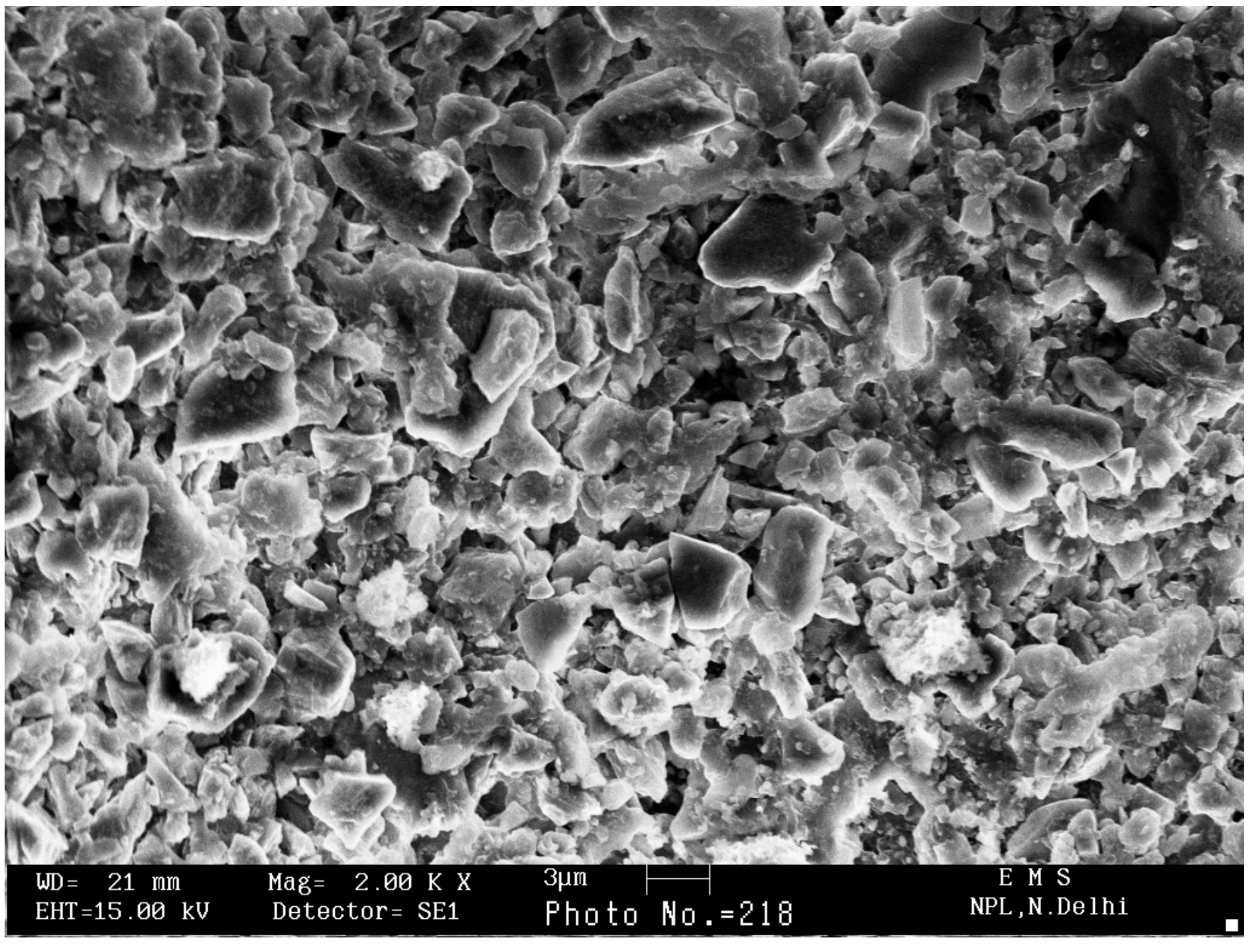

Figure 3(b)

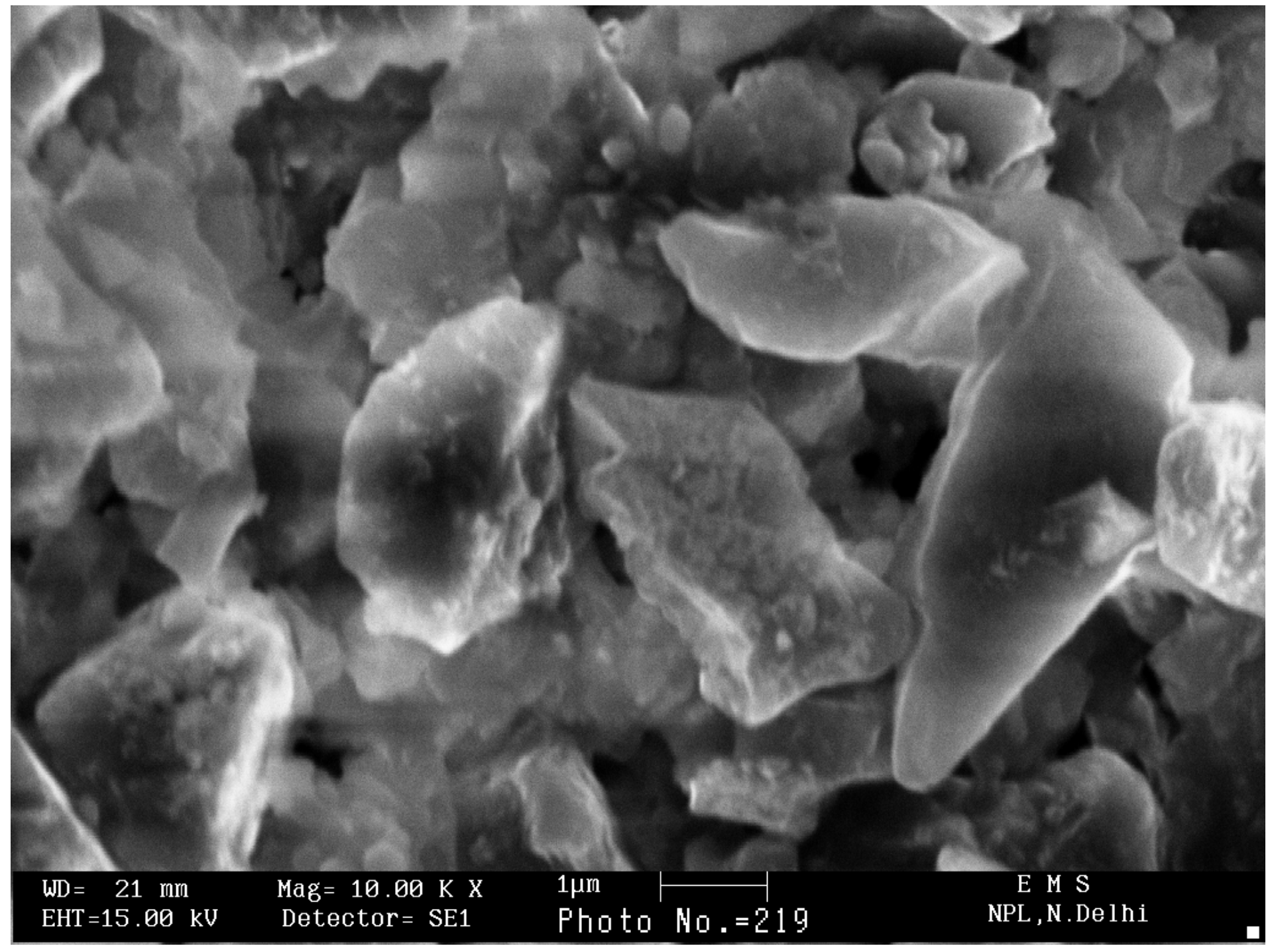

Figure 4 (a)

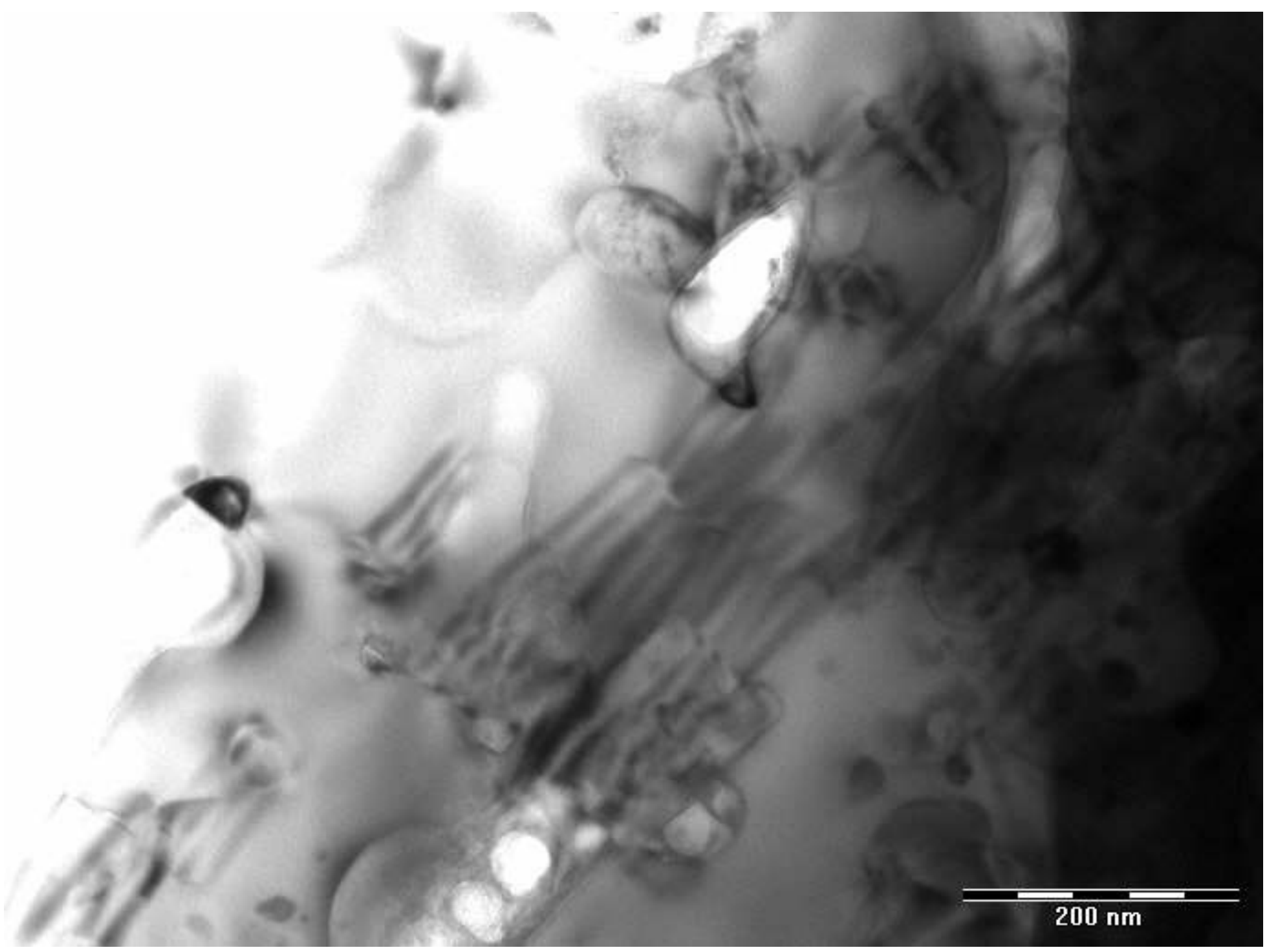

Figure 4 (b)

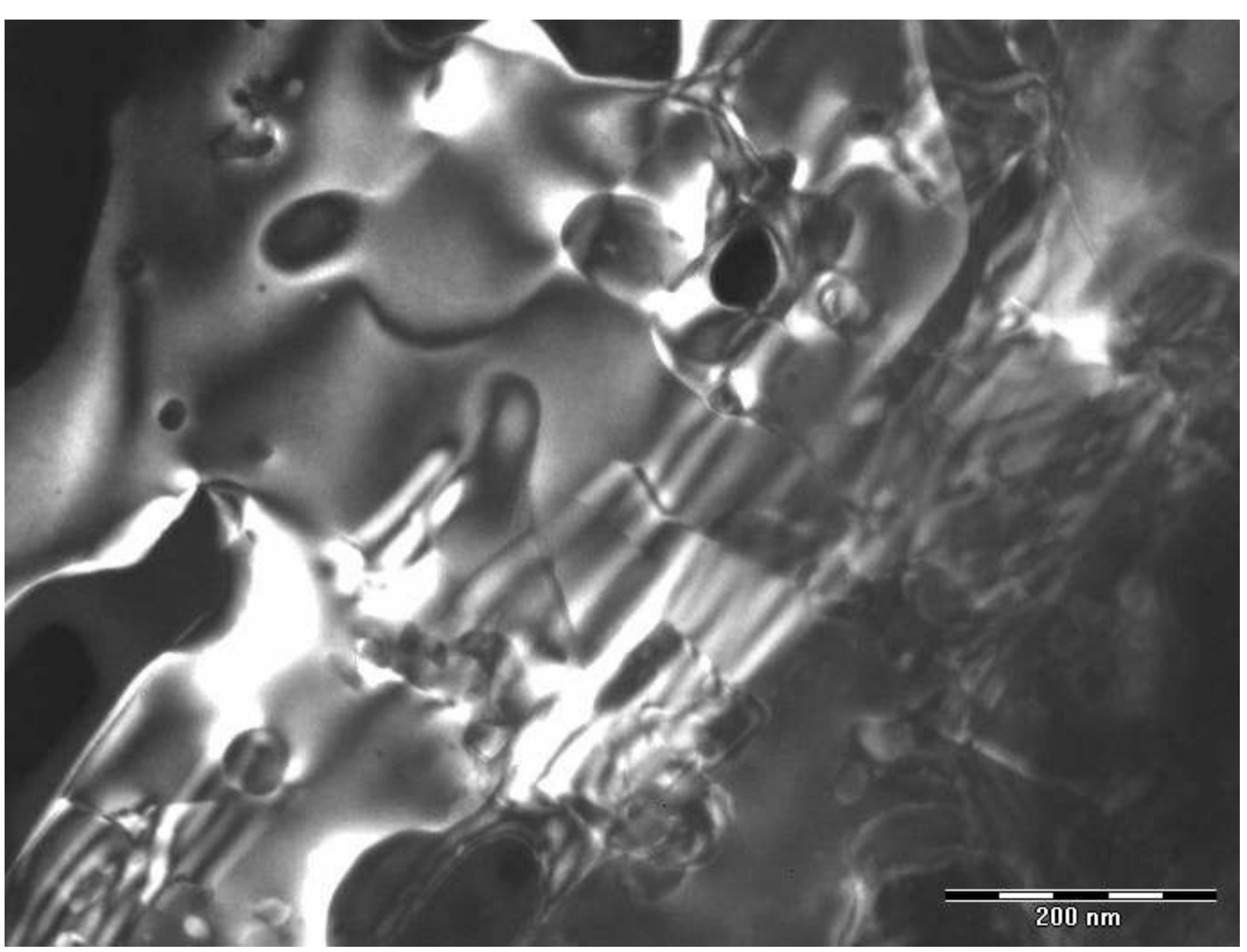

Figure 4 (c)

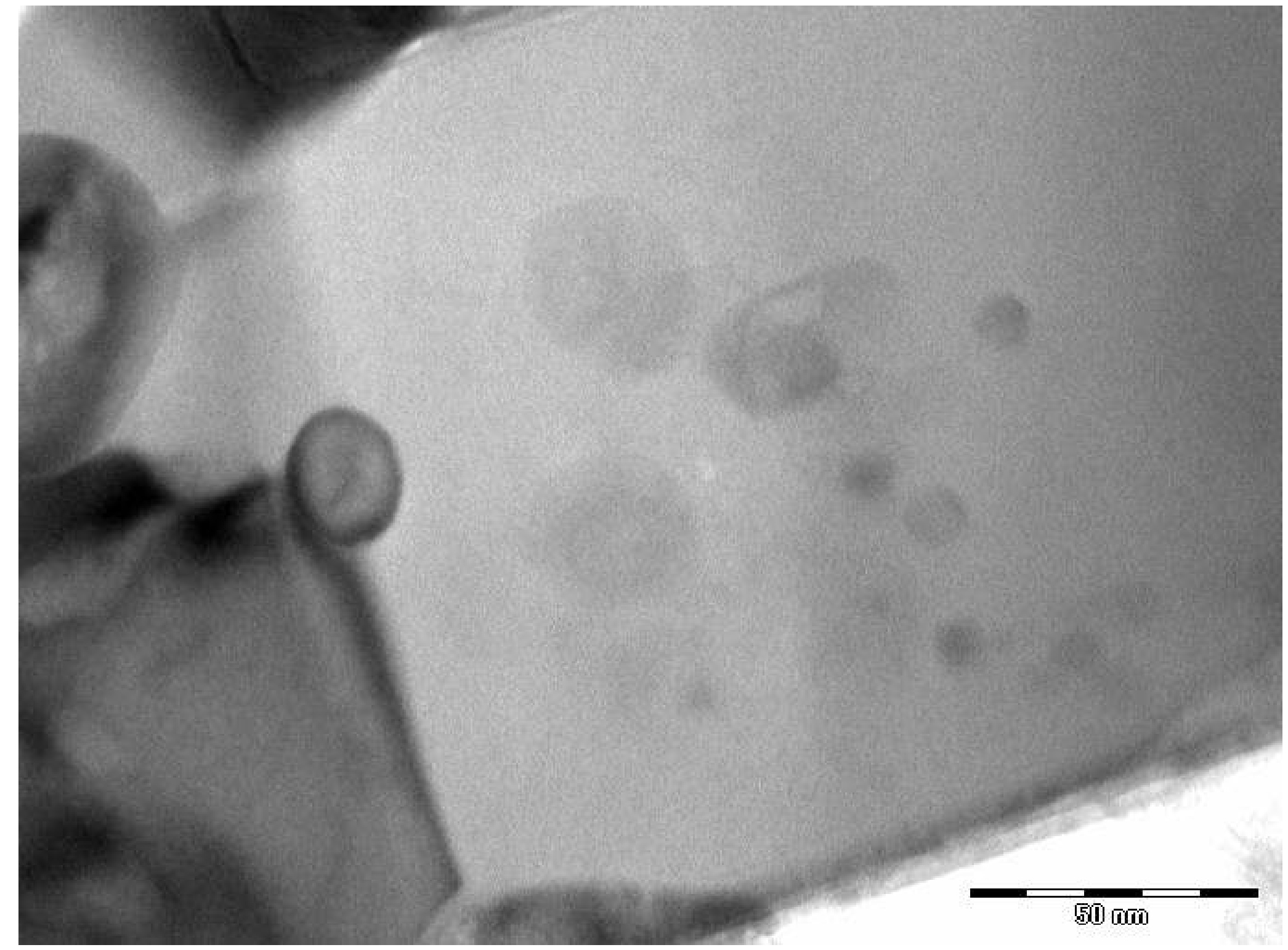

Figure 5(a)

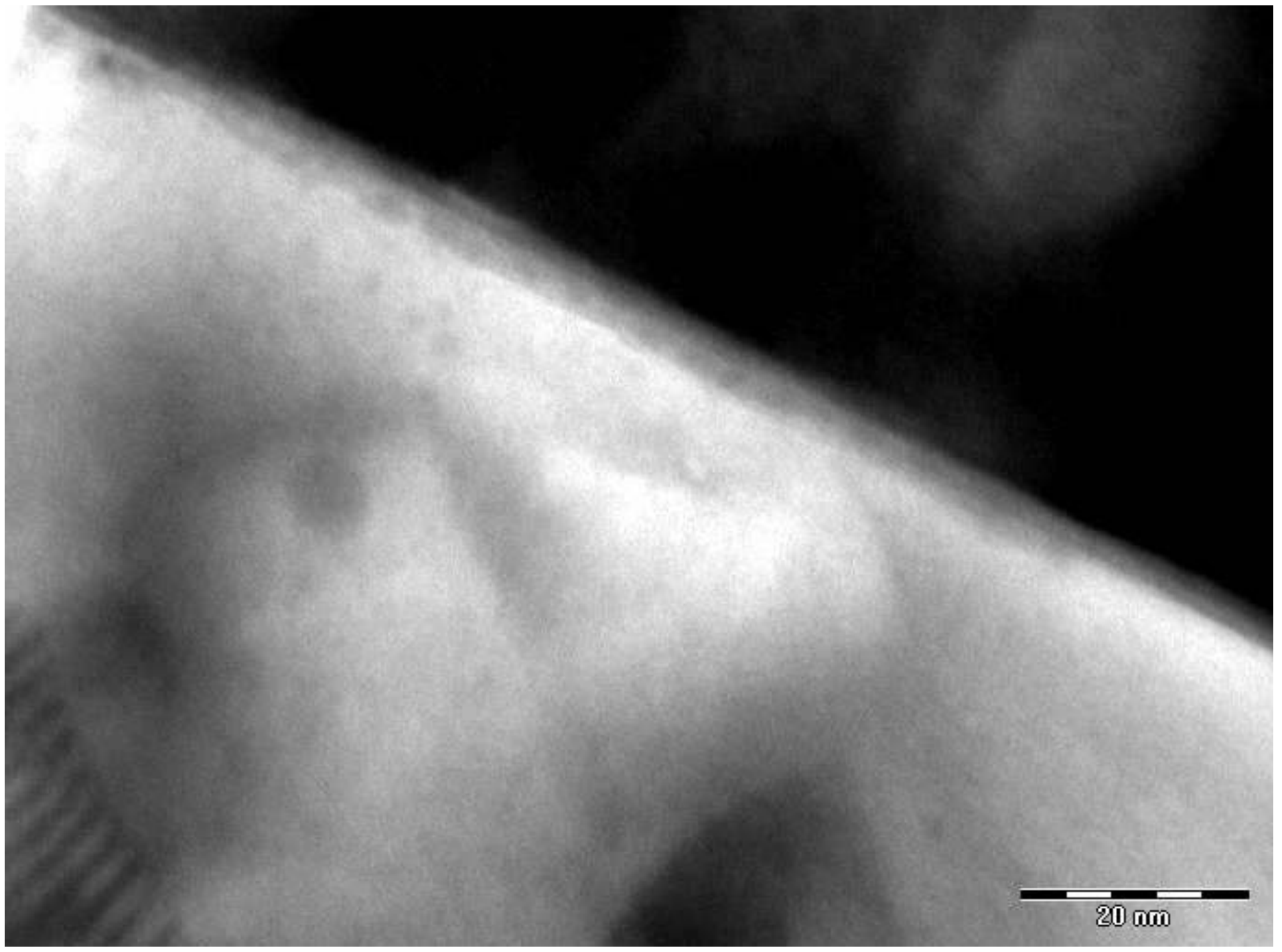

Figure 5(b)

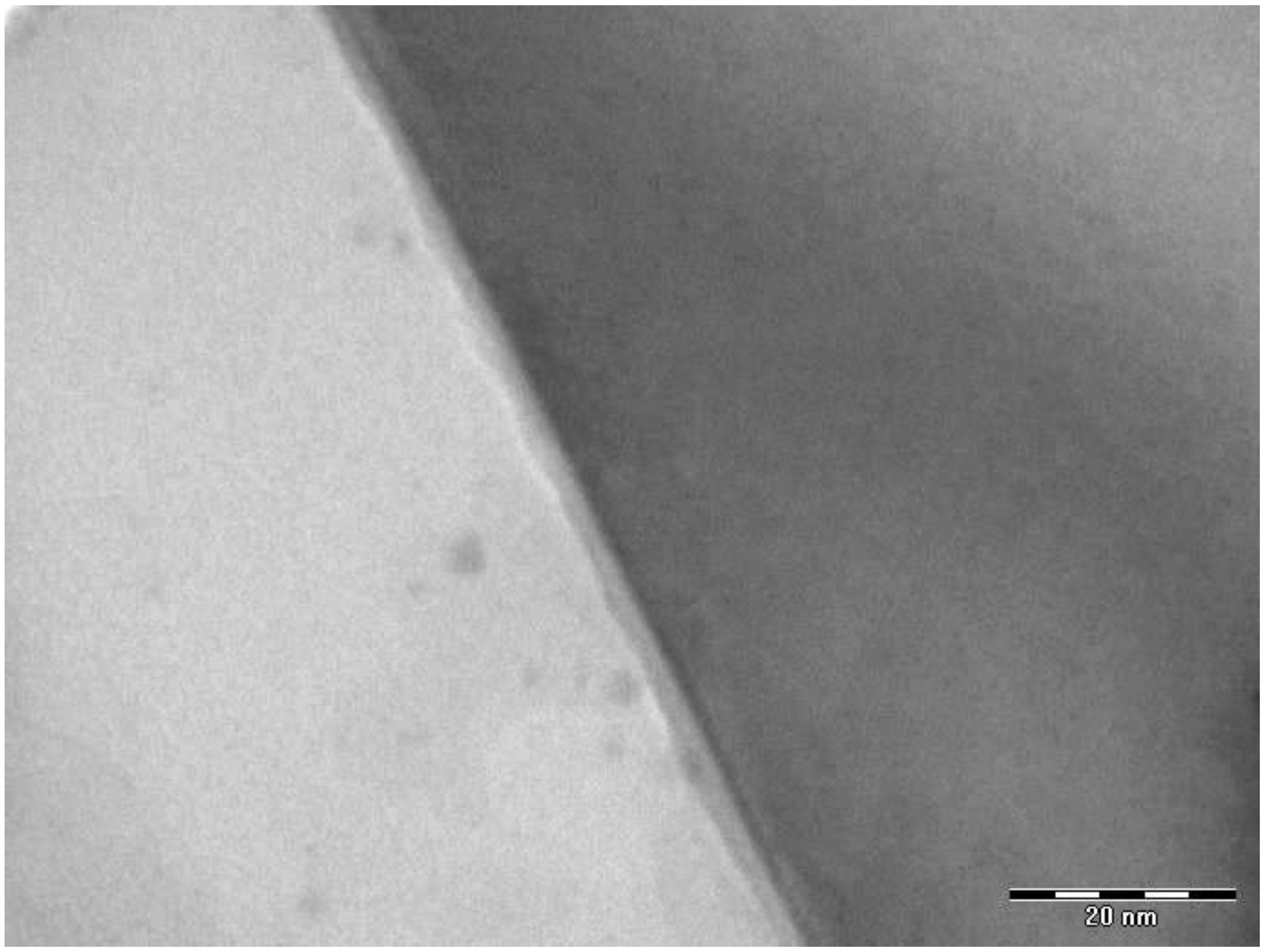

Figure 6

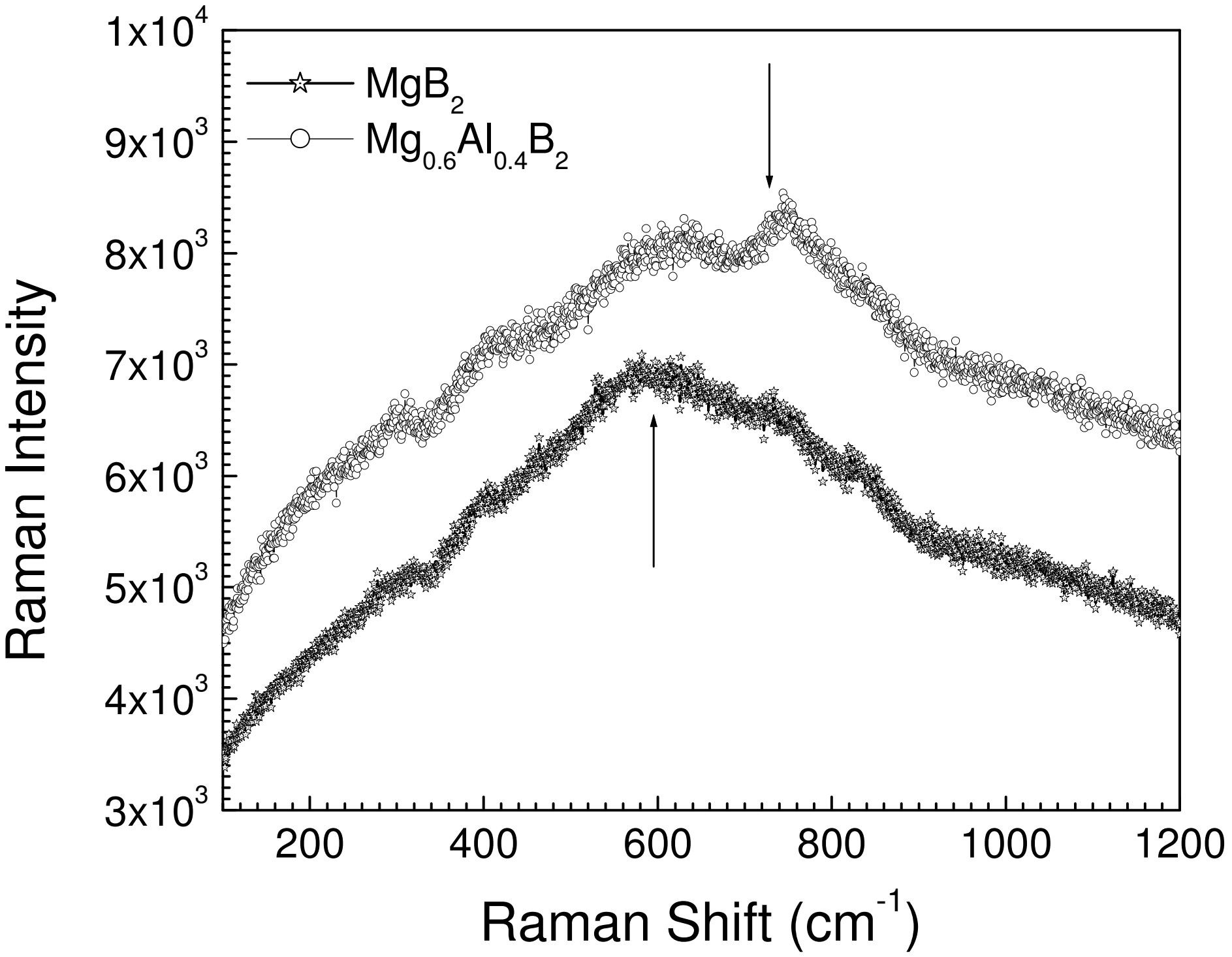

Figure 7

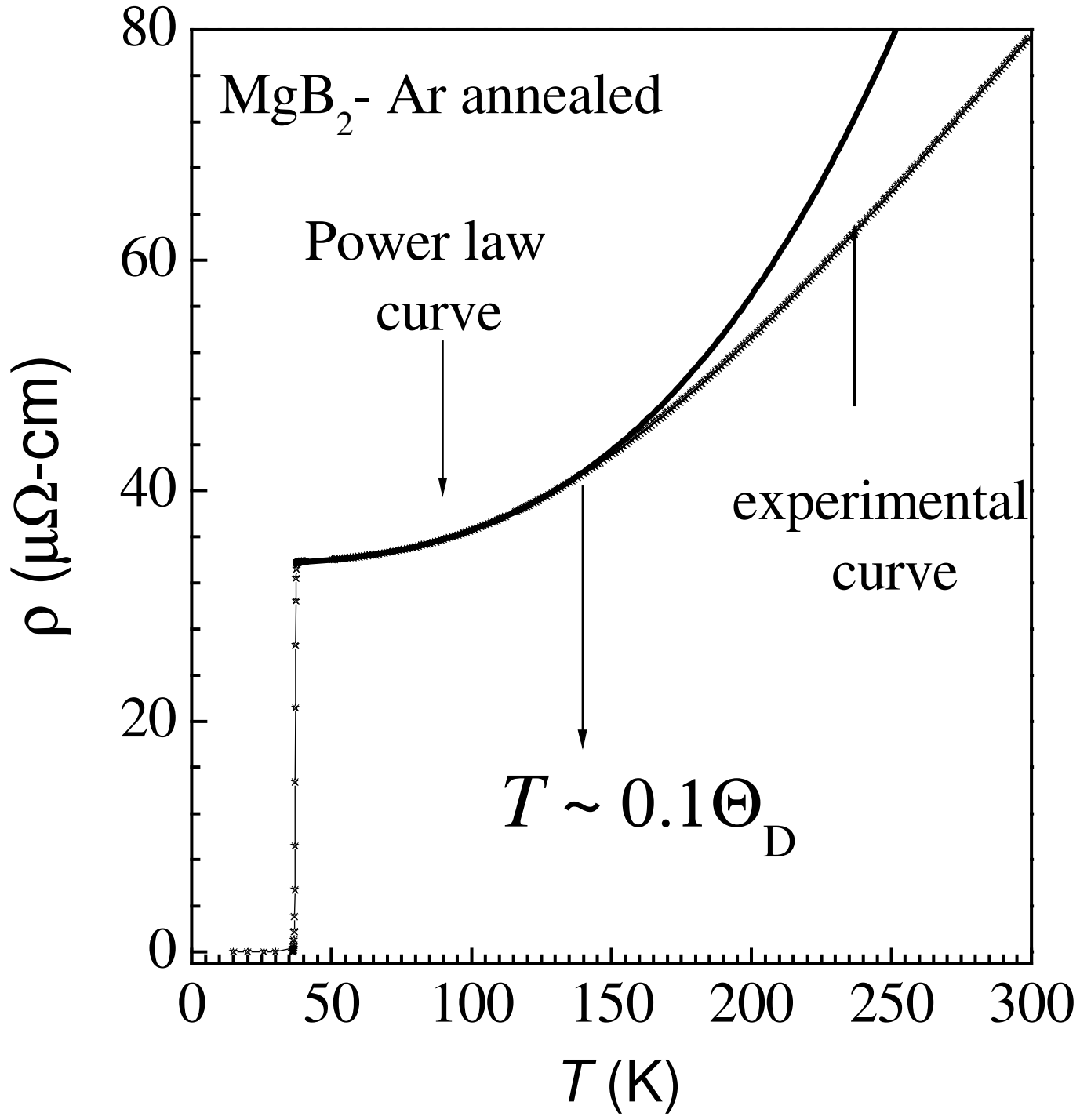

Figure 8

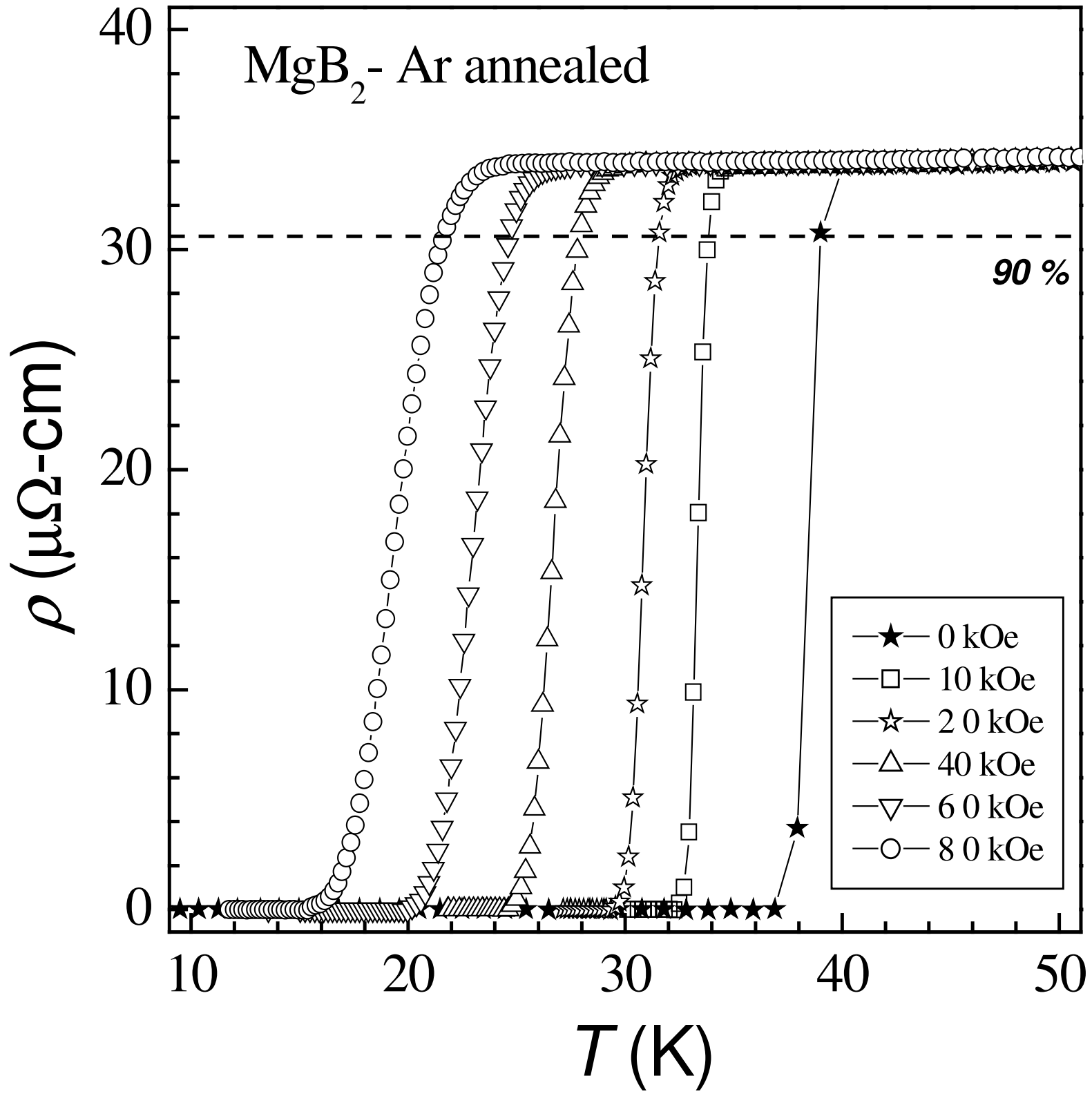

Figure 9

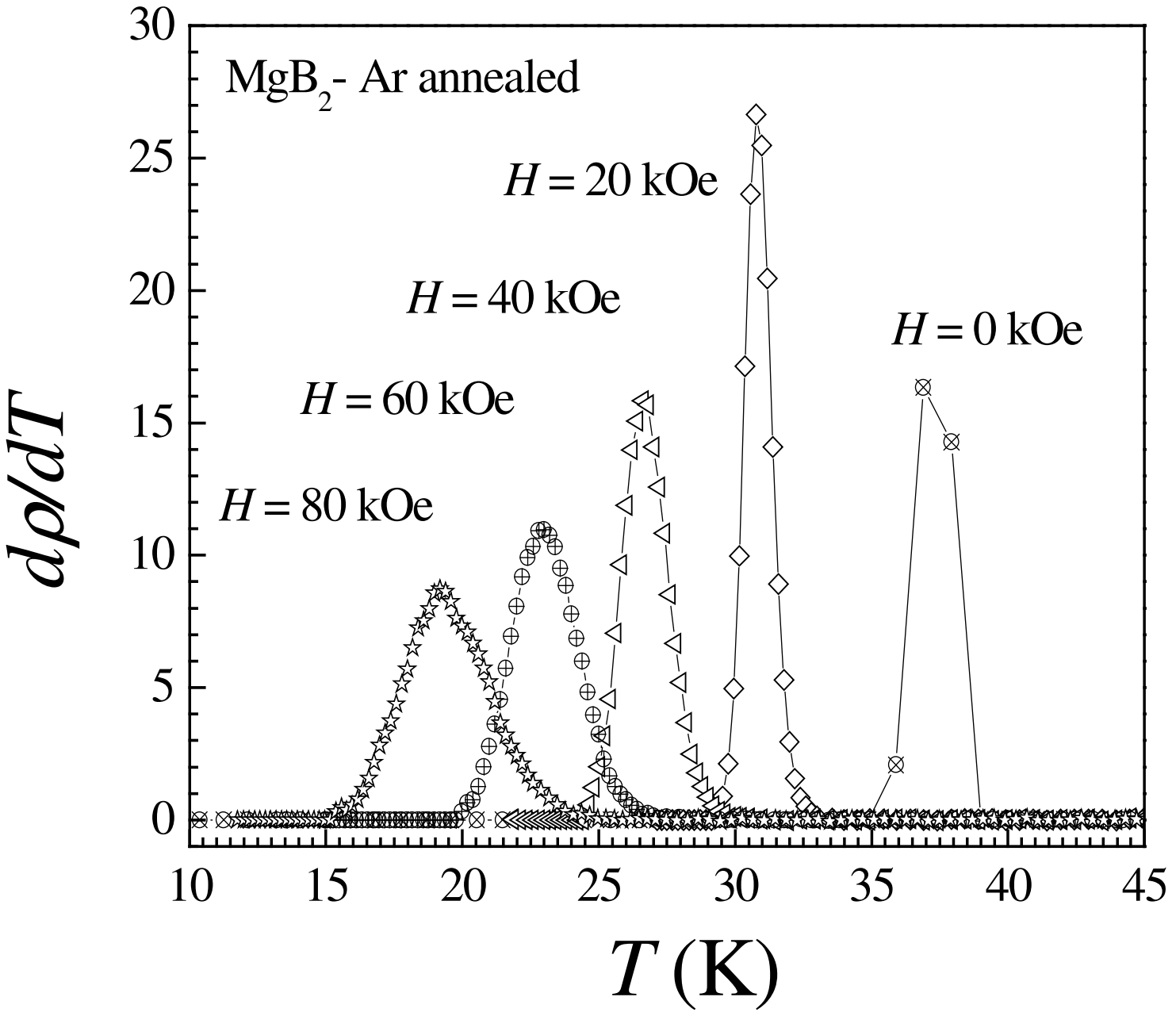

Figure 10

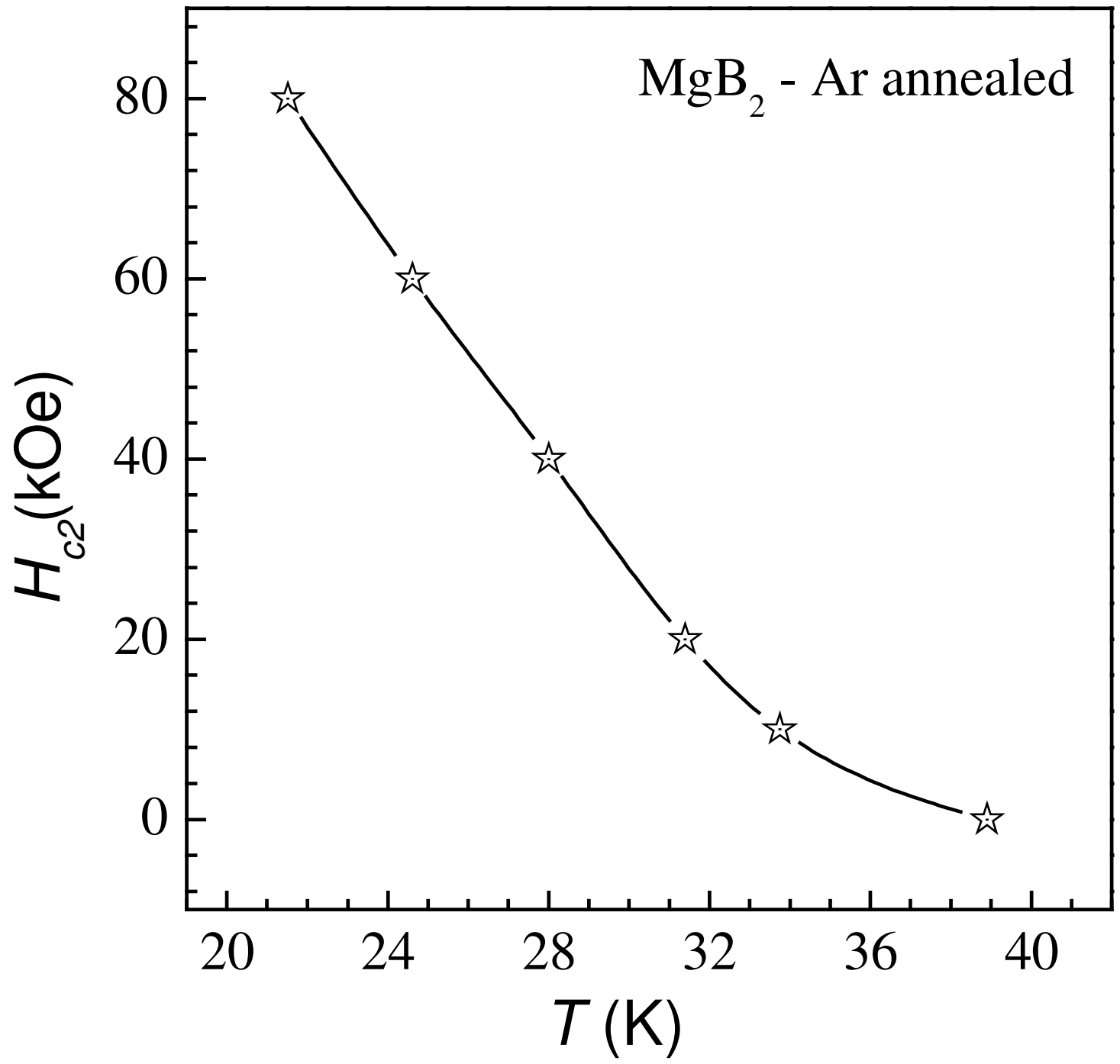

Figure 11

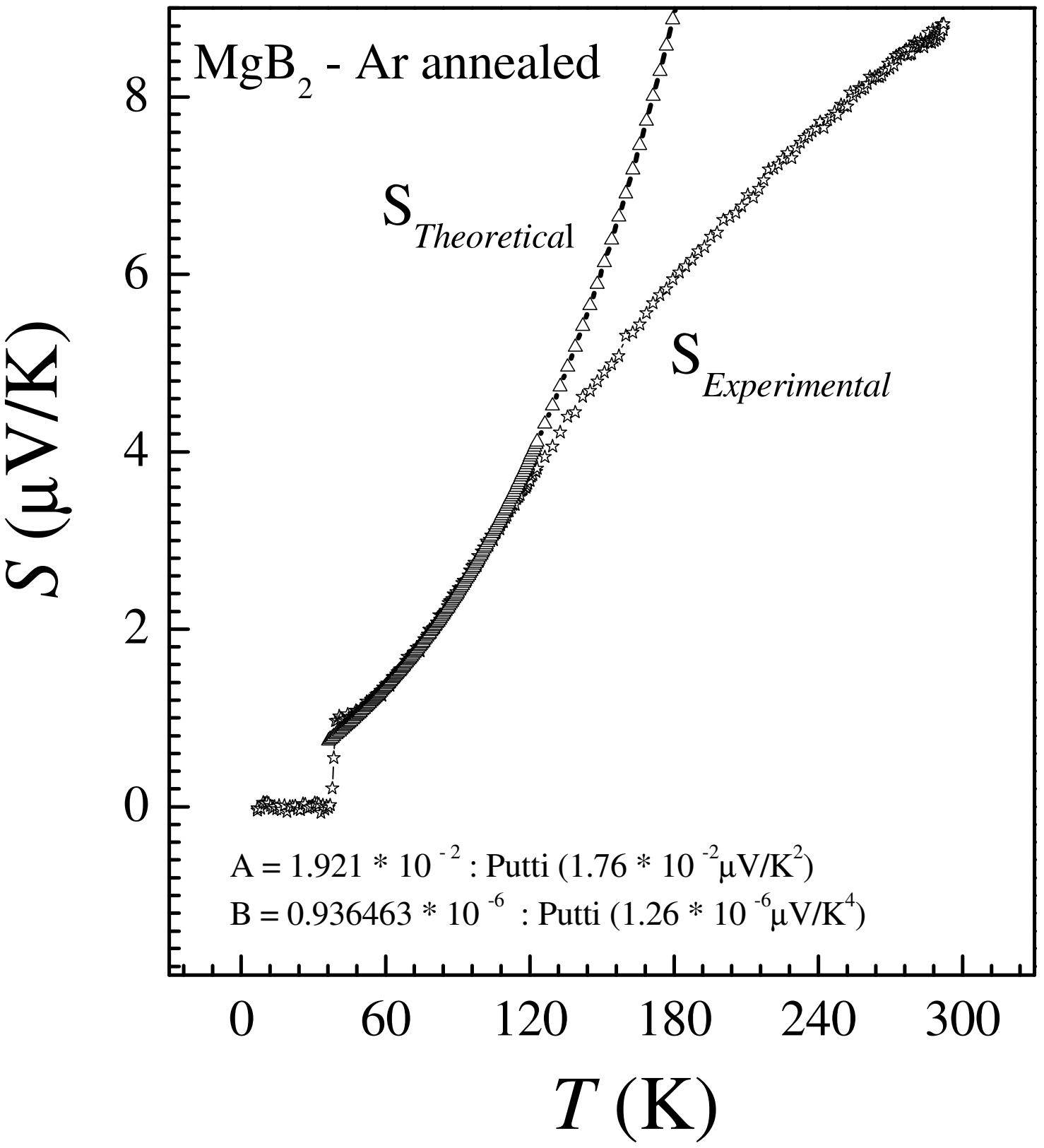

Figure 12

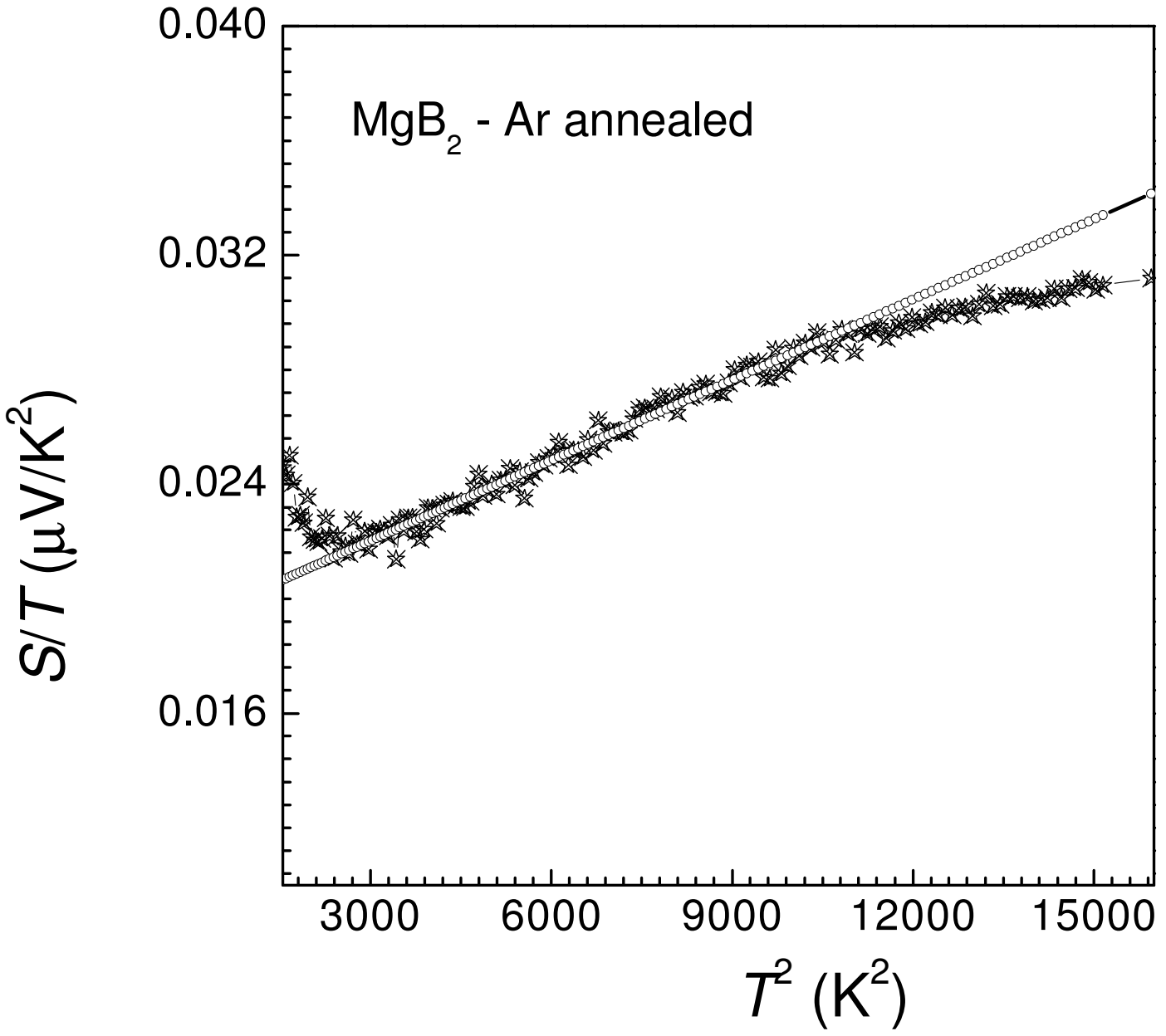

Figure 13

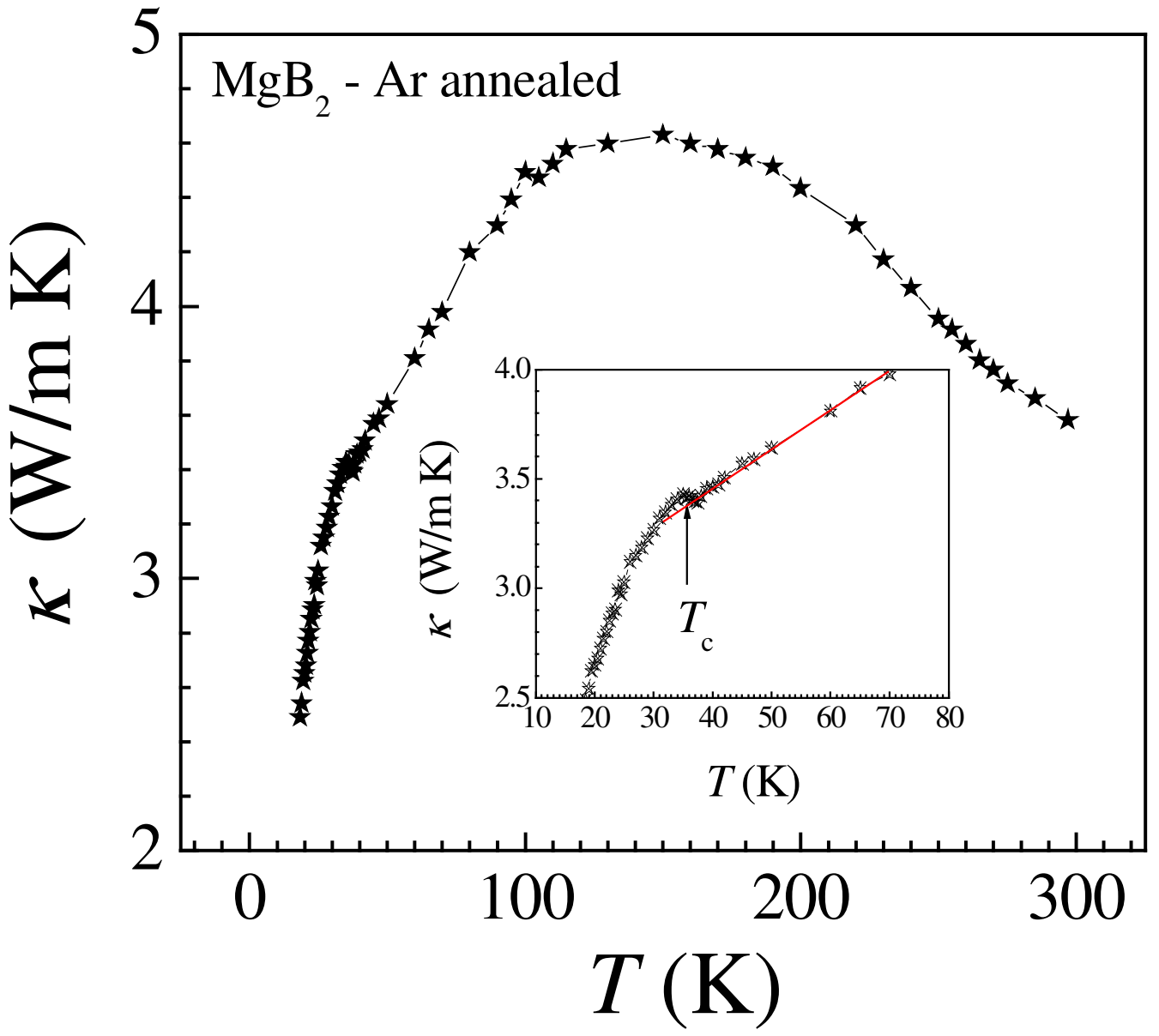

Figure 14

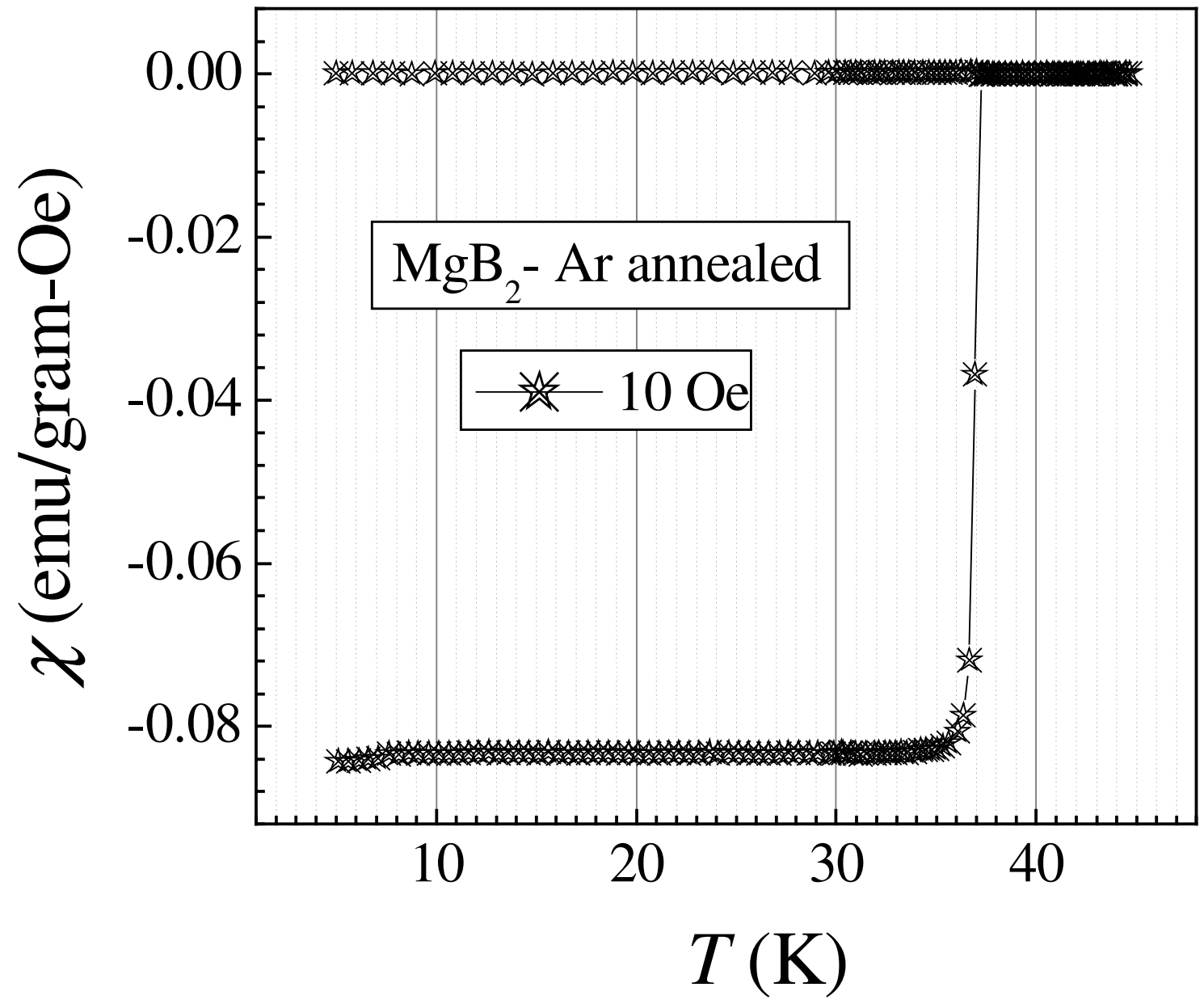

Figure 15

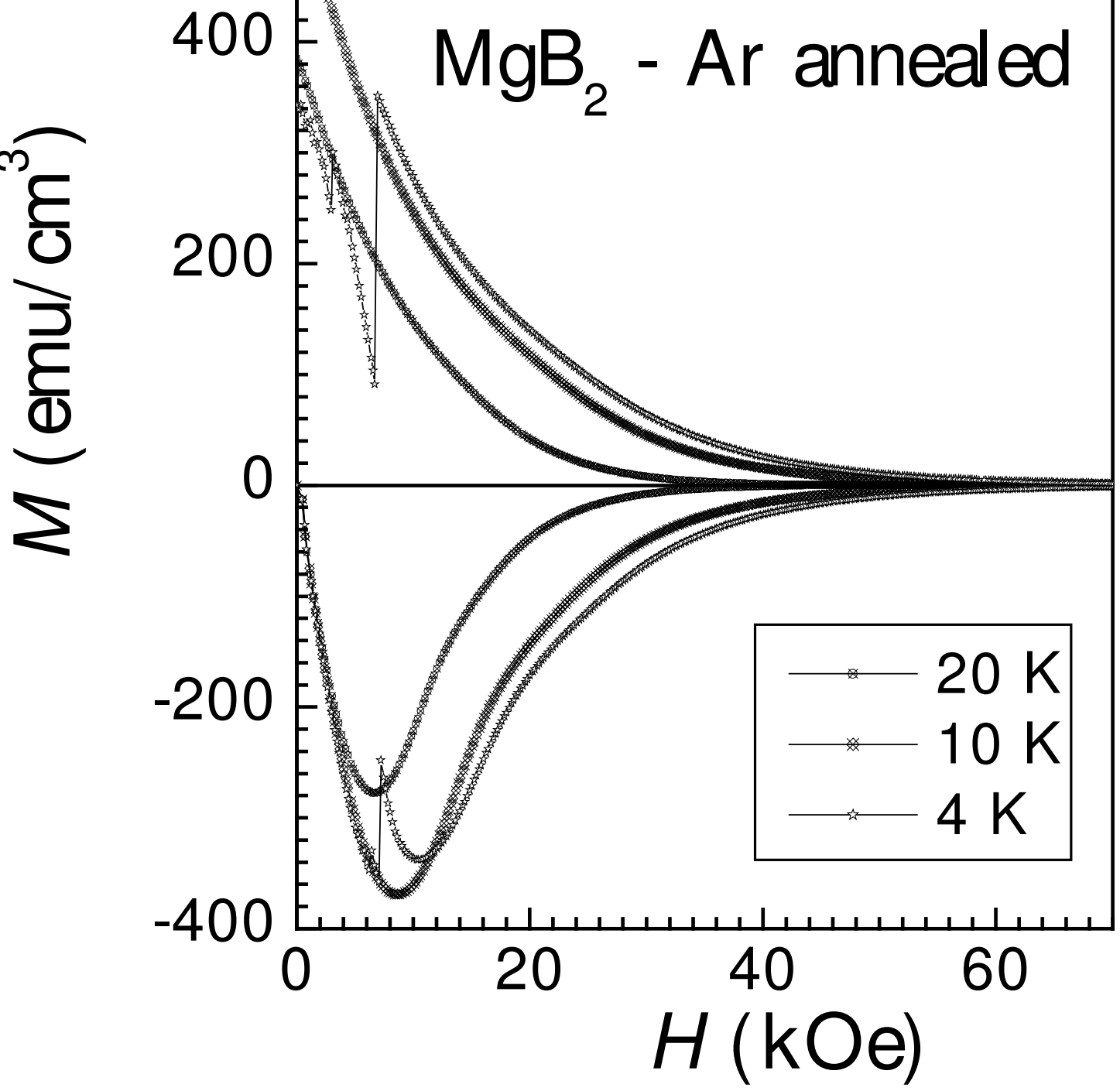

Figure 16

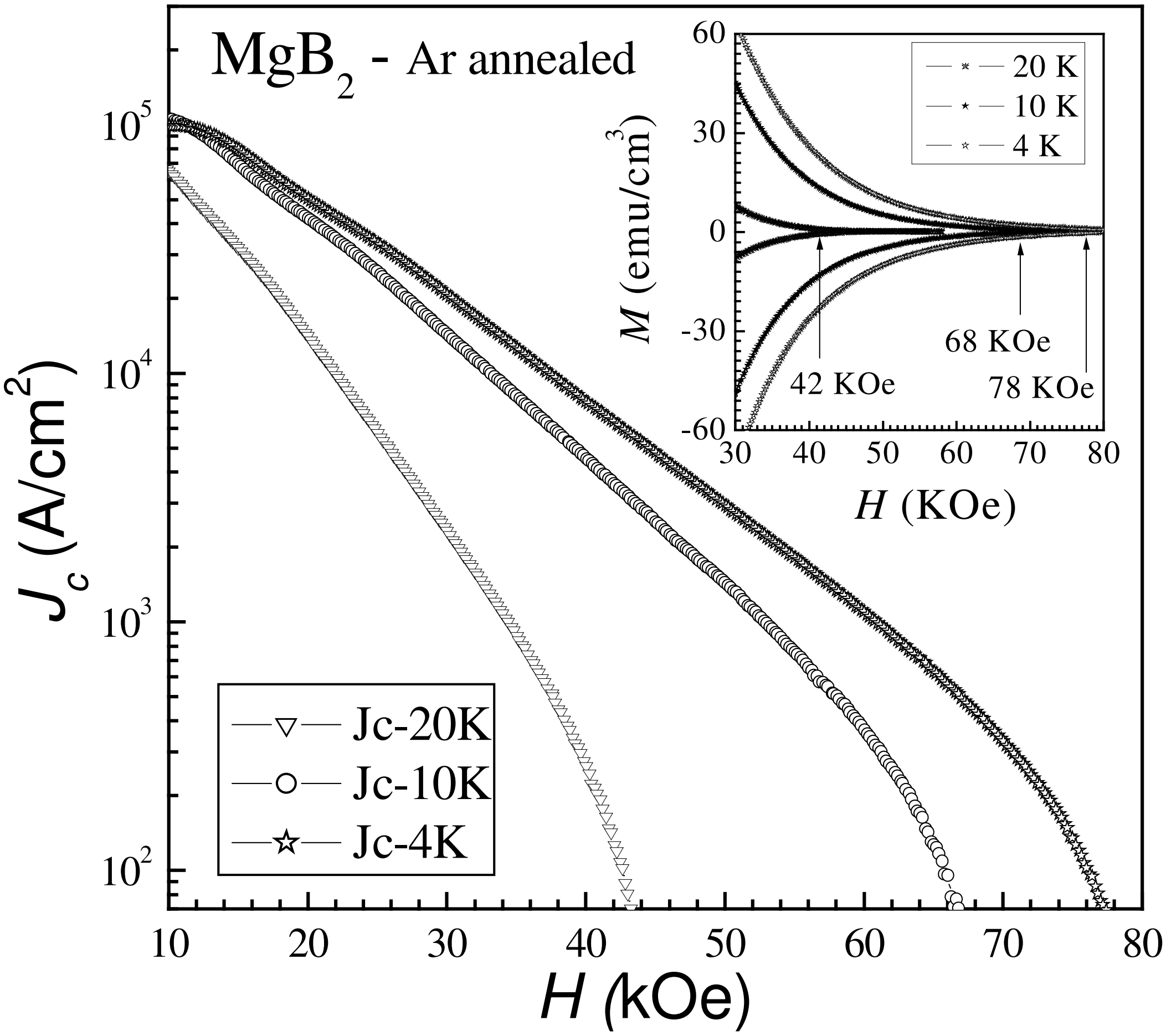

Figure 17

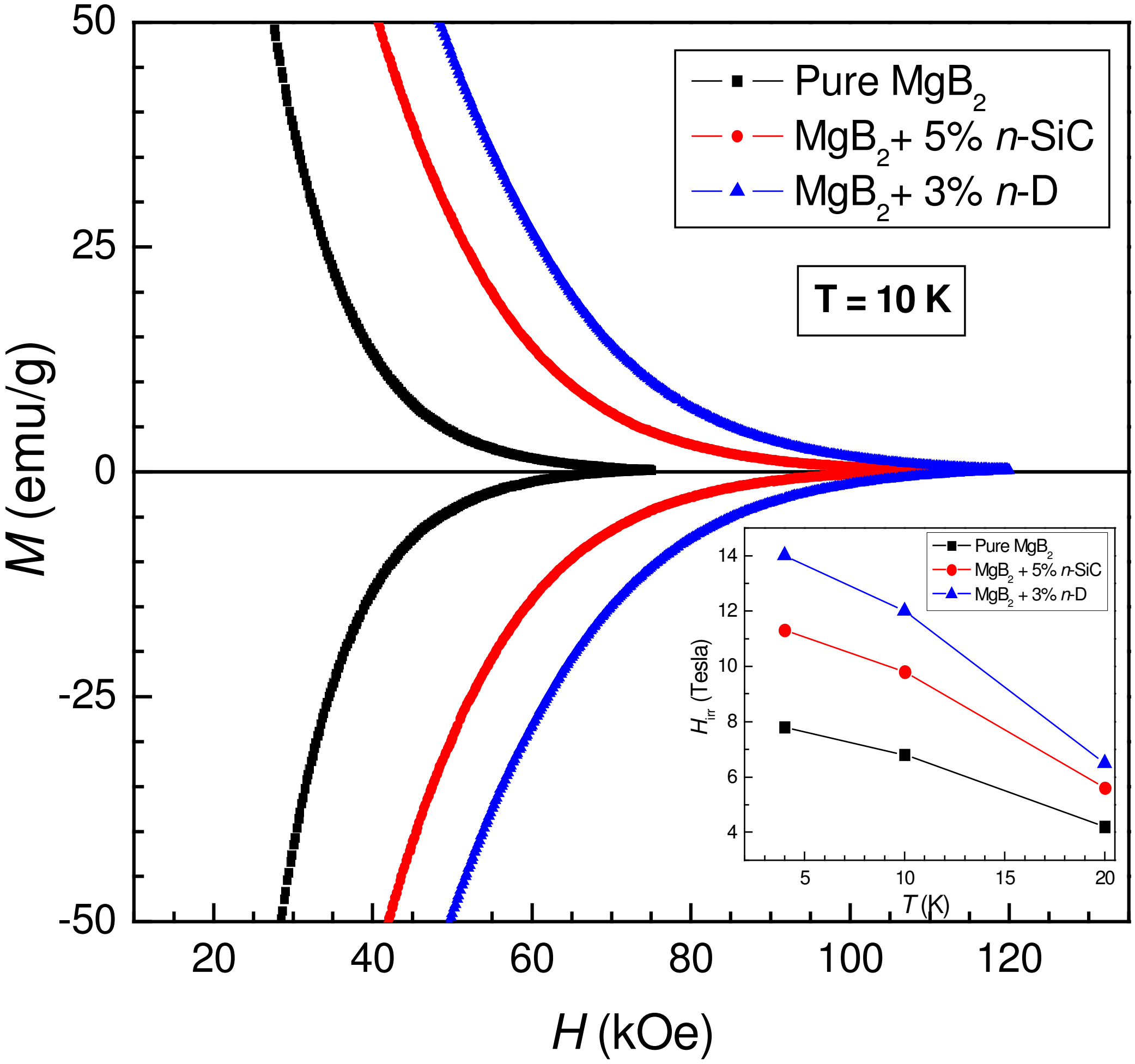

Figure 18

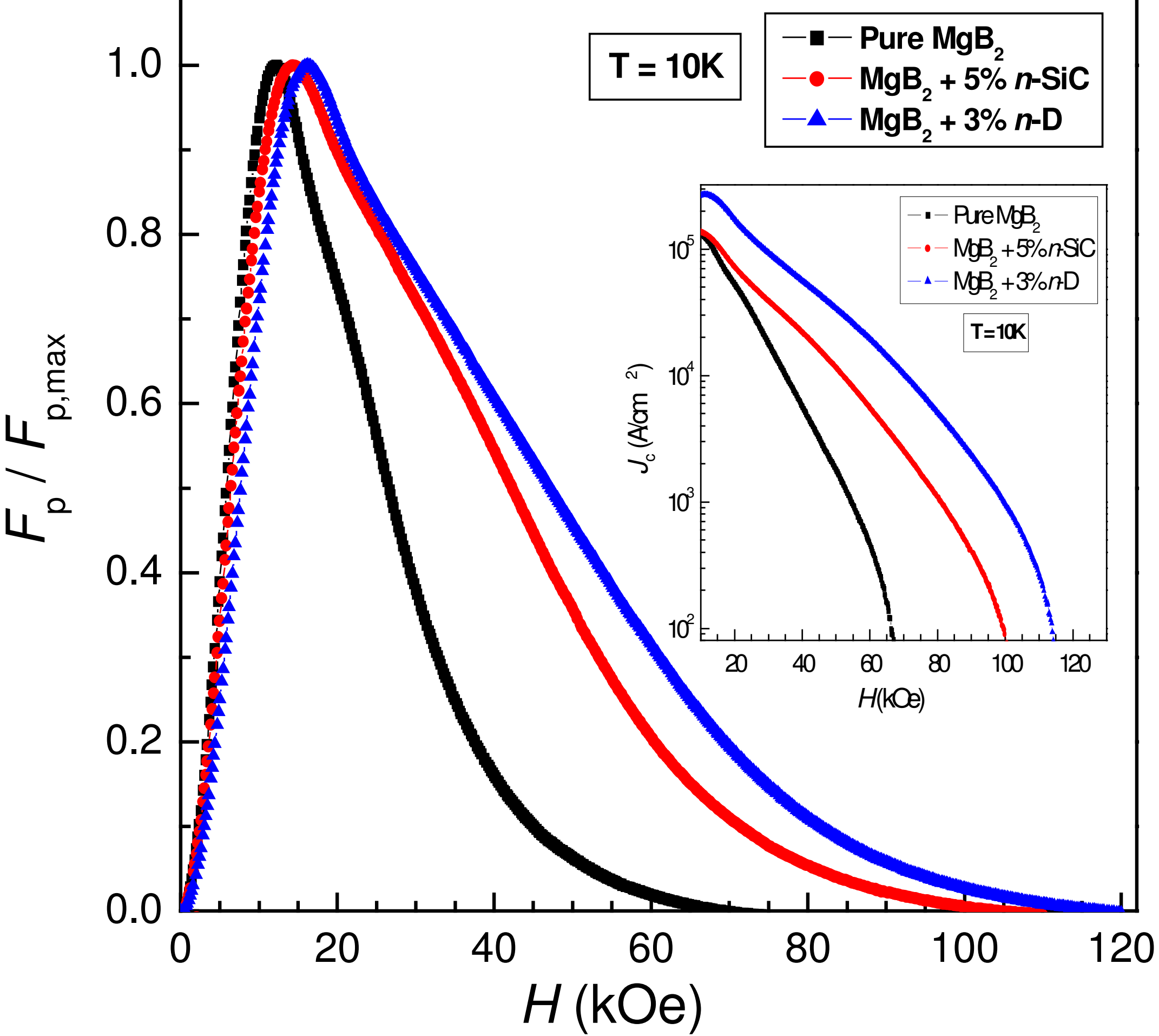